\documentclass[aps,prb,english,twocolumn,showpacs,preprintnumbers,amsmath,amssymb,floatfix,superscriptaddress]{revtex4-1}

\usepackage[T1]{fontenc}
\usepackage[latin9]{inputenc}
\usepackage{pstricks,pstricks-add} 
\usepackage{graphicx}
\usepackage{amssymb}
\usepackage{amsmath}
\usepackage{bm}
\usepackage{relsize}
\usepackage{babel}
\usepackage{verbatim}
\usepackage[colorlinks=true,urlcolor=blue,linkcolor=red,citecolor=red]{hyperref} %%hyperref inserted by Shajesh for providing hyperlinks.
\makeatletter

\usepackage[version=3]{mhchem}

\usepackage{color}

\usepackage{ulem}

\newcommand{\tcc}[1]{\textcolor{cyan}{#1}}%% For using cyan in the text

\makeatother
\usepackage{babel}
\makeatother

\begin{document}
\title{Effect of excess charge carriers and fluid medium on the magnitude and the sign of the Casimir-Lifshitz torque}

\author{P. Thiyam}
%\email{priyadarshini.thiyam@teokem.lu.se}
\affiliation{Division of Theoretical Chemistry, Lund University, P.O. Box 124, SE-221 00 Lund, Sweden}

\author{Prachi Parashar}
%\email{prachi.parashar@ntnu.no}
\affiliation{John A. Logan College, Carterville, IL 62918 USA}
\affiliation{Department of Energy and Process Engineering, Norwegian University of Science and Technology, NO-7491 Trondheim, Norway}

\author{K. V. Shajesh}
%\email{kvshajesh@gmail.com}
\affiliation{Department of Physics, Southern Illinois University-Carbondale, Carbondale, IL 62901 USA}
\affiliation{Department of Energy and Process Engineering, Norwegian University of Science and Technology, NO-7491 Trondheim, Norway}

\author{O. I. Malyi}
%\email{oleksandrmalyi@gmail.com}
\affiliation{Centre for Materials Science and Nanotechnology, University of Oslo, P. O. Box 1048 Blindern, NO-0316 Oslo, Norway}
%\affiliation{School of Materials Science and Engineering, Nanyang Technological University, 50 Nanyang Avenue, Singapore 639798, Singapore}

\author{M. Bostr{\"o}m}
%\email{Mathias.A.Bostrom@ntnu.no}
\affiliation{Department of Energy and Process Engineering, Norwegian University of Science and Technology, NO-7491 Trondheim, Norway}
\affiliation{Centre for Materials Science and Nanotechnology, University of Oslo, P. O. Box 1048 Blindern, NO-0316 Oslo, Norway}

\author{K. A. Milton}
%\email{kmilton@ou.edu}
\affiliation{Homer L. Dodge Department of Physics and Astronomy, University of Oklahoma, Norman, Oklahoma 73019, USA}

\author{I. Brevik}
%\email{iver.h.brevik@ntnu.no}
\affiliation{Department of Energy and Process Engineering, Norwegian University of Science and Technology, NO-7491 Trondheim, Norway}

\author{J. Forsman}
%\email{jan.forsman@teokem.lu.se}
\affiliation{Division of Theoretical Chemistry, Lund University, P.O. Box 124, SE-221 00 Lund, Sweden}

\author{C. Persson}
%\email{clas.persson@fys.uio.no}
\affiliation{Centre for Materials Science and Nanotechnology, University of Oslo, P. O. Box 1048 Blindern, NO-0316 Oslo, Norway}
\affiliation{Department of Materials Science and Engineering, Royal Institute of Technology, SE-100 44 Stockholm, Sweden}
%\affiliation{Department of Physics, University of Oslo, P. O. Box 1048 Blindern, NO-0316 Oslo, Norway}

\vspace{2mm}
\date{\tcc{September 23, 2019}}

\begin{abstract}

Last year, we reported a perturbative theory of the Casimir-Lifshitz torque between planar biaxially anisotropic materials in the retarded limit [Phys. Rev. Lett. {\bf 120}, 131601 (2018)], which is applied here to study the change of sign and magnitude of the torque with separation distance in biaxial black phosphorus having excess charge carriers. The study is carried out both in vacuum as well as in a background fluid medium. The presence of extra charge carriers and that of an intervening fluid medium are both found to promote enhancement of the magnitude of the torque between identical slabs. The degree of enhancement of the magnitude of torque increases not only with an increased carrier concentration but also with separation distance. In the non-identical case when different planes of anisotropic black phosphorus face each other, owing to the non-monotonic characteristic of the sign-reversal effect of the torque, the enhancement by carrier addition and intervening medium also becomes non-monotonic with distance.
%as compared to the torque in the identical orientation of the carrier-free black P slabs. 
In the presence of a background medium, the non-monotonic degree of enhancement of the torque with distance is observed even between identical slabs. 

\end{abstract}

%Anisotropy, interaction energy, perturbative method, biaxial, Phosphorene, DFT

%\pacs{34.20.Cf; 42.50.Lc; 92.20.Uv}
\maketitle

%-----------------------------------------------
\section{Introduction}

The introduction of a pair of material slabs in vacuum breaks its translational symmetry giving rise to the Casimir force between the slabs. Birefringence or anisotropy in the material properties of the slabs further induces breaking of the rotational symmetry of the vacuum in which case a torque appears between the slabs tending to rotate and align their principal optical axes~\cite{DLP}. Parsegian and Weiss presented an analytical calculation of this torque in the non-retarded limit between semi-infinite uniaxial bulk slabs interacting across an anisotropic medium~\cite{ParsegianWeiss1972}, while Barash derived the same in the retarded limit but in an isotropic medium~\cite{Barash1978}. Last year~\cite{PRL2018}, we demonstrated, using a perturbative theory, that the retarded Casimir torque between two biaxially polarizable material slabs can exhibit a non-trivial reversal of sign as a function of separation distance if the in-planar frequency-dependent polarizability components of one of the dielectric slabs intersect at a certain frequency. In the uniaxial limit, this theory yields the exact non-retarded result of Barash while for the retarded case, the results were found to match numerically for reasonable values of the perturbative parameter~\cite{PRL2018}. Very recently, Broer {\it et al.} reported an independent derivation via Maxwell eigenmode approach confirming Barash's results~\cite{Broer2019}.

Though the Casimir-Lifshitz torque was theoretically predicted as early as 1961~\cite{DLP}, it was only in 2018 that its existence was experimentally confirmed~\cite{SomersNature2018}. With this accomplishment, further investigations of the magnitude and nature of the torque are indispensable for the ultimate goal of feasible practical applications. The smallness of the magnitude of torque has so far thwarted experimental progress.
The magnitude of the torque for a pair of discs each of cross-sectional area 1 ${\mu}$m$^2$ at a separation distance of micrometer range is consistently reported in the literature to be of the order of 10$^{-20}$ Nm for several uniaxial materials~\cite{MundayPRA2005, Iannuzzi2006,HBChan}. While this degree of sensitivity has now been established as achievable by the precision experiment in Ref.~\cite{SomersNature2018}, theoretical efforts to augment the magnitude of torque from material considerations~\cite{MundayPRA2005, Iannuzzi2006} as well as techniques such as application of an external magnetic field~\cite{Sirvent2010} had been underway. Reference~\cite{Lindel2018} predicted a Casimir torque between photonic topological insulators tunable by an external magnetic field. Reference~\cite{SomersPRL2018} showed that an intervening dielectric medium, contrary to the expectation that the medium would screen the interaction, could instead enhance the magnitude of torque by a factor of 2. More recently, Ref.~\cite{Antezza2019} reported a giant Casimir torque between rotated finite gratings, that is several times larger in magnitude than between the corresponding infinite ones. 
Optical modulation of the Casimir force by varying density of charge-carriers had been demonstrated in Refs.~\cite{ChenPRL2006, Chen2007}.

%%%%%%%%%%%%%%%%%%%%%%%%

In this work, we shall demonstrate an increase in the magnitude of the Casimir-Lifshitz torque between a pair of biaxial slabs as a result of carrier insertion as well as the presence of an intervening medium between the two slabs. We shall also show non-trivial changes of the sign of the torque with separation distance~\cite{PRL2018} for these cases, while elucidating why this particular effect fails to show up in the experiment of Ref.~\cite{SomersNature2018}. We consider biaxially anisotropic bulk black phosphorus (black P) as an illustration. This material displays a unique puckered honeycomb structure rendering a non-planar feature to the layer. Its exhibition of in-plane anisotropy of various material properties has been attributed to this feature~\cite{LiuACSnano2014}. The layered bulk crystal is of the van der Waals type but with a more sophisticated nature of interlayer interaction~\cite{ShulenburgerNanolett2015} exhibiting varying band gap energies inversely proportional to the number of layers~\cite{WangNatNano2015, LiNatNano2016,Maruyama1981}. 
Moreover, it has high carrier mobility~\cite{LiuACSnano2014, QiaoNatComm2014, YueLiu1, YueLiu2}, high reactivity~\cite{Preactivity1, Preactivity2, Poxidation1, Poxidation2} and high exciton binding energy~\cite{Phighexcitonbe, WangNatNano2015} making it a desirable material for potential device applications.
We explore the possibility of modifying the anisotropy through addition of free electrons (which can either be achieved by dopants or by carrier injection) where the mobility of the additional free carriers is governed by the dispersion of the conduction band. In the following section, we briefly present the perturbative method of Ref.~\cite{PRL2018} for computing the torque, generalised to the case when the intervening medium is an isotropic dielectric medium. In section III, we describe the model of carrier injection in bulk black P, for which the electronic structure and the dielectric tensor were computed by density functional theory (DFT). {\tcc {It should be noted that, while the theory presented here can handle finite thicknesses of the interacting slabs, we are considering only black P slabs of infinite thicknesses. This is necessitated by the fact that the procedure of carrier injection requires bulk structures.}}  The analysis of the torque with the resulting dielectric tensors is presented in section IV, and the effect of the presence of an intervening medium is discussed in section V. In section VI, we present a brief section on the enhancement featuring cases with large magnitudes of the torque, and we end with conclusions in section VII. We also provide an appendix elaborating the expressions for the reduced reflection coefficients of the retarded interaction between biaxial planar materials of finite thickness, and another on the anisotropic structure of black P.
 
%----------------------------------------------
\section{Perturbative formalism for Casimir-Lifshitz torque}

%****************************************************

\begin{figure}
\includegraphics[width=8cm]{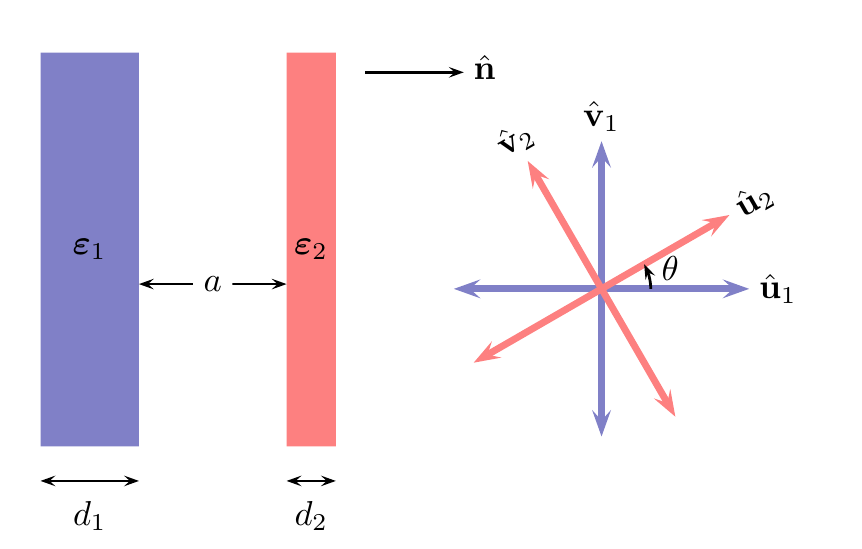}%
\caption{Two biaxial dielectric slabs, of thicknesses $d_i$,
with dielectric functions described by Eqs.\,(\ref{aniso-diel}),
separated by distance $a$. The orientation of the principal
axes $\hat{\bf u}_1$ and $\hat{\bf v}_1$ of the first slab
relative to the principal axis of the second slab
$\hat{\bf u}_2$ and $\hat{\bf v}_2$ is illustrated on the right.
The third principal axes, $\hat{\bf n}$, for both the slabs is chosen normal
to the plane of the slabs. {\tcc {However, in this work, we are considering only black P slabs of infinite thicknesses in order to facilitate carrier addition.}} }
\label{anisotropic-slabs-fig}
\end{figure}

Let us consider two biaxial dielectric slabs, of thicknesses $d_1$ and $d_2$,  as shown in Fig.~\ref{anisotropic-slabs-fig},
with principal axes of their respective dielectric tensors along
($\hat{\bf u}_1$, $\hat{\bf v}_1$, $\hat{\bf n}$)
and
($\hat{\bf u}_2$, $\hat{\bf v}_2$, $\hat{\bf n}$),
\begin{subequations}
\begin{eqnarray}
{\bm\varepsilon}_1(\omega) =
\varepsilon^{u}_1(\omega) \,\hat{\bf u}_1 \hat{\bf u}_1
+ \varepsilon^{v}_1(\omega) \,\hat{\bf v}_1 \hat{\bf v}_1
+ \varepsilon^{n}_1(\omega) \,\hat{\bf n} \hat{\bf n}, \\
{\bm\varepsilon}_2(\omega) =
\varepsilon^{u}_2(\omega) \,\hat{\bf u}_2 \hat{\bf u}_2
+ \varepsilon^{v}_2(\omega) \,\hat{\bf v}_2 \hat{\bf v}_2
+ \varepsilon^{n}_2(\omega) \,\hat{\bf n} \hat{\bf n},
\end{eqnarray} 
\label{aniso-diel}%
\end{subequations}%
%----------
%----------
where we have chosen one of the principal axes of the slabs
to be along $\hat{\bf n}$.
Further, we choose the slabs to be parallel and their
normals to be in the direction of $\hat{\bf n}$.
The distance between the inner surfaces of the slabs is $a$. 
The intervening medium is assumed to be isotropic, characterized by the frequency-dependent dielectric function, $\varepsilon_3(\omega)$. In what follows, the frequency dependence is suppressed for brevity.

In the perturbative approach, following the theory developed in Refs.~\cite{gearsI,gearsII,QFEXT2009} for the perturbation in the shape of the material slabs, we choose the decomposition of the 
dielectric function of the biaxial material to be 
\begin{equation}
{\bm\varepsilon}_i = \bar{{\bm\varepsilon}}_i
+ \Delta{\bm\varepsilon}_i
\end{equation}
with $\bar{\bm\varepsilon}_i$ representing a uniaxial 
background with
\begin{equation}
\varepsilon^\perp_i = \frac{\varepsilon^{u}_i + \varepsilon^{v}_i}{2}
\quad \text{and} \quad \varepsilon^{||}_i = \varepsilon^{n}_i.
\label{iso-sep-rdf}
\end{equation}
%using ${\bf 1} = \hat{\bf u}_i \hat{\bf u}_i + \hat{\bf v}_i \hat{\bf v}_i$.
$\varepsilon^{||}_i$ is parallel with the direction normal to the surface.
The degree of 
anisotropy of each material is defined as 
\begin{equation}
\beta_i = \frac{\varepsilon^{u}_i -\varepsilon^{v}_i }
{\varepsilon^{u}_i +\varepsilon^{v}_i },
\label{beta_para}
\end{equation}
which is the perturbation parameter in our theory. 
Thus the anisotropy leading to the biaxial nature of the material is then
completely captured inside
\begin{equation}
\Delta{\bm\varepsilon}_i =
\left(\frac{\varepsilon^{u}_i -\varepsilon^{v}_i}{2}\right) 
\,(\hat{\bf u}_i \hat{\bf u}_i -\hat{\bf v}_i \hat{\bf v}_i), 
\end{equation}%
This particular choice renders
\begin{equation}
\text{Tr} \, {\Delta\bm\varepsilon}_i = 0,
\end{equation}
and ensures that the first order contribution in $\beta_i$ is zero.
The leading-order contribution to the interaction energy per unit area between the two biaxial slabs originates from the second-order terms in the perturbation parameters, i.e. $\beta_1^2$, $\beta_2^2$, and $\beta_1\beta_2$. We are particularly interested in the latter, which depends on the angle $\theta$ between the in-planar principal axes of the two slabs, and gives rise to a nonvanishing torque:

%\begin{subequations}
%\begin{eqnarray}
\begin{equation}
%\begin{align}
%\begin{split}
\mathcal{T}^{(2)} %&= \frac{\partial\mathcal{E}^{(2)}(a,\theta)}{\partial\theta} \\
= \! -\frac{k_BT  \sin 2\theta}{2 \pi }\!
\sum_{m=0}^{\infty}{'}\! \beta_{1}\! \beta_{2}\! \!\int_0^\infty\!\!\!\!\! k\, \mathrm{d}k\,
\textrm{Tr}(\widetilde{\bm R}_1\!\widetilde{\bm R}_2) e^{-2\kappa_3 a},
\label{tor-def}
%\end{split}
%\end{align}
\end{equation}
where $k_B$ is the Boltzmann constant and $T$ the temperature. The prime on the summation denotes that the 
zero frequency mode is taken with half-weight. We define $\kappa_3 = \sqrt{k^2 +\zeta_m^2\varepsilon_3/c^2}$, where $\zeta_m=2\pi mk_BT/\hbar$ is the imaginary Matsubara frequency, $\omega=i\zeta$. The reduced reflection coefficients 
$\widetilde{\bm R}_i$ are defined in Appendix A.

\section{Calculation of the dielectric tensor for black P with excess of free carriers}

We model bulk black P (space group Cmce, number 64) with the primitive cell comprising four P atoms with the electronic configuration [Ne]:3s$^2$p$^3$; thus $n=20$ valence electrons. Addition of excess charge carriers implies that a fraction $dn$ of an electron is introduced in each cell, and the carrier concentration $n_e$ is $dn/V$, where $V$ is the volume of the primitive cell.
We expect this method of carrier insertion to be a reasonable assumption for low-to-moderate doping of
shallow donors or for carrier injection by weak applied electric fields. For doping with shallow donors, the donor electrons form a free electron gas in the conduction band. A proper choice of dopant elements will have negligible impact on the electronic structure of the host material~\cite{Clas1}. For carrier-injection, the electrons diffuse into the material by the electric field and form an excess of free carriers. Alternatively, or in combination, the free carriers can be achieved by optical excitation from absorption of light. For weak applied electric fields, the electronic structure of the host material will not be greatly affected~\cite{Clas2}. Here, we will not consider alterations of the host material due to electric field or dopants, but explore how an electron gas in the conduction band affects the dielectric functions and subsequently the torque.

With the free electrons in the conduction band, a plasma contribution to the response function appears in the system~\cite{Mahan,Ambrosch-Draxl2006}. We consider some values of fraction $dn$ that give carrier concentration $n_e$ in the range of $1\times$10$^{19}$\textendash$5\times$10$^{20}$ cm$^{-3}$, with the expectation that a concentration of the order of 10$^{20}$ cm$^{-3}$ is the very upper region that can be achieved with only small impact on the host material~\cite{Pref}.

We determine the dielectric tensors of phosphorous for different carrier concentration $n_e$ from electronic structure calculations using VASP (Vienna ab initio simulation package)~\cite{VASP1, VASP2, VASP3}. The bulk P structure is relaxed with the optB88-vdW functional~\cite{Klimes1, Klimes2} to treat the exchange-correlation part of the layered configuration, with the cut-off energy for the plane-wave basis set fixed at 600 eV. The resulting volume of the unit cell is 80.1 {\AA$^3$}. That yields fractions 0.001, 0.01 and 0.04 as some of the suitable choices of $dn$. The resulting number of carriers $n_e$ injected per cm$^3$ and the corresponding plasma frequencies~\cite{Ambrosch-Draxl2006} along the three principal directions are presented in Table~\ref{dN}. The converged plasma frequencies are determined by calculating the intraband transitions in the lowest conduction band using the modified Becke-Johnson (mBJ) functional~\cite{mBJ1, mBJ2} with a cut-off energy of 325 eV for the plane-wave basis set. To ensure convergence, a large, automatic $k$-mesh of $50\times50\times54$ is used for the Brillouin zone summation with a Gaussian smearing of 0.02 eV. 
From the plasma frequency the dielectric response due to the free carriers is determined with the Drude model assuming a damping constant of 0.2 eV~\cite{DrudeConstant}. In addition, the dielectric response from the host, carrier-free material is determined by calculating the optical interband transitions using the Heyd-Scuseria-Ernzerhof (HSE06) functional~\cite{HSE03, HSE06} and a $8\times8\times10$ $k$-mesh. {\tcc {These data for black P without extra charge carriers calculated using DFT are comparable to the available low-frequency experimental data~\cite{BPexpt1,BPexpt2,BPexpt3}.}}

\begin{table}
\begin{center}
\caption{The fraction $dn$ of electron added per unit cell, the resulting concentration of extra carriers $n_e$ per cm$^3$ and the generated plasma frequencies along the three principal directions. }
%\centering
\small\addtolength{\tabcolsep}{4pt}
\label{dN}
\vspace{5px}
%\begin{adjustbox}{max width=\textwidth}
\begin{tabular}{ccccc}
\hline
\hline
\multicolumn{1}{c}{$dn$} &\multicolumn{1}{c}{$n_e$ (cm$^{-3}$)} &\multicolumn{3}{c}{plasma frequency (eV)} \\
\cline{3-5}
& & \multicolumn{1}{c}{$xx$} & \multicolumn{1}{c}{$yy$} &\multicolumn{1}{c}{$zz$} \\
\hline
0.001      &  $1.25 \times$10$^{19}$     & 0.11 & 0.28 & 0.31 \\
 \hline
0.01      & $1.25\times$10$^{20}$   & 0.42 & 0.70 & 0.91 \\
 \hline
0.04      & $4.99 \times$10$^{20}$          & 0.98 & 1.37 & 1.42 \\
\hline
\hline
\end{tabular}
%\end{adjustbox}
\end{center}
\end{table}

\begin{widetext}
\hspace{7px}
\begin{figure}[h]
\centering
\includegraphics[width=13cm]{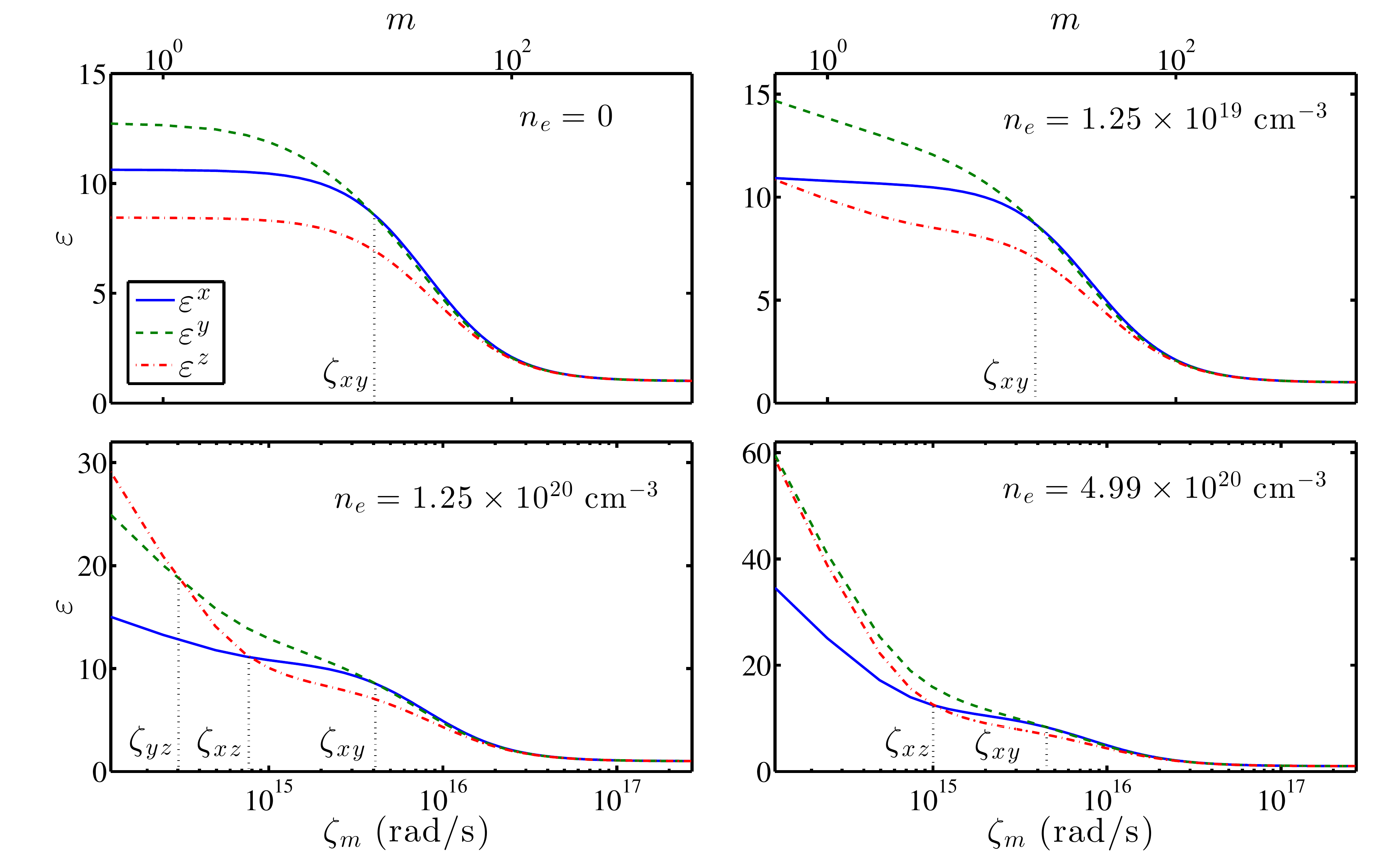}
\caption{(Color online) The principal components $\varepsilon^x$, $\varepsilon^y$ and $\varepsilon^z$ of the dielectric tensors at room temperature of carrier-free black P and carrier-excess black P with different concentrations of extra electrons $n_e$. The Matsubara frequency $\zeta_m$ of the bottom $x$-axis and the Matsubara number $m$ of the top $x$-axis correspond to each other. The dielectric values at $m=0$ are placed on the $y$-axis. $\varepsilon^x$ and $\varepsilon^y$ form the planar components while $\varepsilon^z$ is the component normal to the face of the P slab. That is, $\varepsilon^x$ is measured along the direction defined by lattice constant 3.34 {\AA}, $\varepsilon^y$ along that defined by lattice constant 4.47 {\AA} and $\varepsilon^z$ along 10.74 {\AA}. See Appendix B for the structure of bulk black P.}
\label{diel_bulkci}
\end{figure}
\end{widetext}

The computed dielectric tensors of the original bulk and the carrier-added bulk P are shown in Fig.~\ref{diel_bulkci}. $\varepsilon^x$ and $\varepsilon^y$ are the planar components that correspond to the principal axes $\hat {\bf u}$ and $\hat {\bf v}$ respectively while $\varepsilon^z$ forms the normal component perpendicular to the slab and corresponds to the principal axis $\hat {\bf n}$ (see Appendix B for the structure of bulk black P and determination of principal axes). The immediate effect seen as a result of carrier insertion is an increase of dielectric values at lower Matsubara frequencies due to intraband transitions. As the concentration $n_e$ of extra carriers increases, the range of frequencies affected increases. {\tcc {Consequently, the intersection of the planar dielectric components $\epsilon^x$ and $\epsilon^y$ occurs at higher frequencies, denoted by $\zeta_{xy}$.}}
%Consequently, the in-planar crossing of $\varepsilon^x$ and $\varepsilon^y$ components, denoted by $\zeta_{xy}$, is pushed towards higher frequencies. 
An additional feature that is observed is the intersection of the $\varepsilon^z$ and $\varepsilon^x$ components that is not seen in the original carrier-free bulk P. As can be noted from Table~\ref{dN}, the plasma frequency along the principal $z$-direction has relatively higher values than in the $x$-direction resulting in higher dielectric values along $z$, notably at lower frequencies. As a result, the static value of $\varepsilon^z$ component becomes nearly equal to that of $\varepsilon^x$ component for $n_e=1.25\times 10^{19}$ cm$^{-3}$. For higher carrier insertion, $\varepsilon^z$ gradually overtakes and crosses the $\varepsilon^x$ component. The plasma frequency along $z$ is also higher than along $y$, and $\varepsilon^z$ overtakes and intersects with $\varepsilon^y$ as well for the case of $n_e=1.25\times 10^{20}$ cm$^{-3}$. The frequencies at which the in-planar dielectric components intersect for the different faces ($XY$, $YZ$ and $XZ$) are tabulated in Table~\ref{inplanar_crossing_freqs} for the computed dielectric tensors.

\begin{table}
\begin{center}
\caption{The frequencies of intersection of the planar dielectric components of the $XY$ plane $\zeta_{xy}$, the $YZ$ plane $\zeta_{yz}$ and the $XZ$ plane $\zeta_{xz}$ in the original P bulk and the carrier-excess P dielectric tensors in rad/s.}
%\centering
\small\addtolength{\tabcolsep}{5pt}
\label{inplanar_crossing_freqs}
\vspace{5px}
%\begin{adjustbox}{max width=\textwidth}
\def\arraystretch{1.4}
\begin{tabular}{cccc}
\hline
\hline
\multicolumn{1}{c}{$n_e$ (cm$^{-3}$)} &\multicolumn{1}{c}{$\zeta_{xy}$} &\multicolumn{1}{c}{$\zeta_{yz}$} &\multicolumn{1}{c}{$\zeta_{xz}$}  \\
\hline
0        & $3.89 \times$10$^{15}$ &-  &-  \\
               
\hline
$1.25 \times$10$^{19}$        & 3.94$ \times$10$^{15}$ & - &-  \\
                
 \hline
$1.25 \times$10$^{20}$      & $4.13 \times$10$^{15}$ & $3.29 \times$10$^{14}$  & $7.9 \times$10$^{14}$  \\

 \hline
$4.99 \times$10$^{20}$          & $4.50 \times$10$^{15}$ &-   & $1.03 \times$10$^{15}$\\
                 
\hline
\hline
\end{tabular}
%\end{adjustbox}
\end{center}
\end{table}

\section{Casimir-Lifshitz torque}
The rich characteristics of the carrier-excess dielectric functions with several in-planar crossings enable a comparative study of the magnitude as well as a detailed analysis of the sign-reversal behaviour of Casimir-Lifshitz torque as a function of distance that we earlier predicted in Ref.~\cite{PRL2018}. In this work, we consider only black P slabs of infinite thicknesses whose in-planar principal axes are at an angle $\theta=\pi/4$, which corresponds to the maximum value of the torque.

\subsection{Identical configuration}

%--------------
\begin{figure}
\begin{center}
\includegraphics[width=9cm]{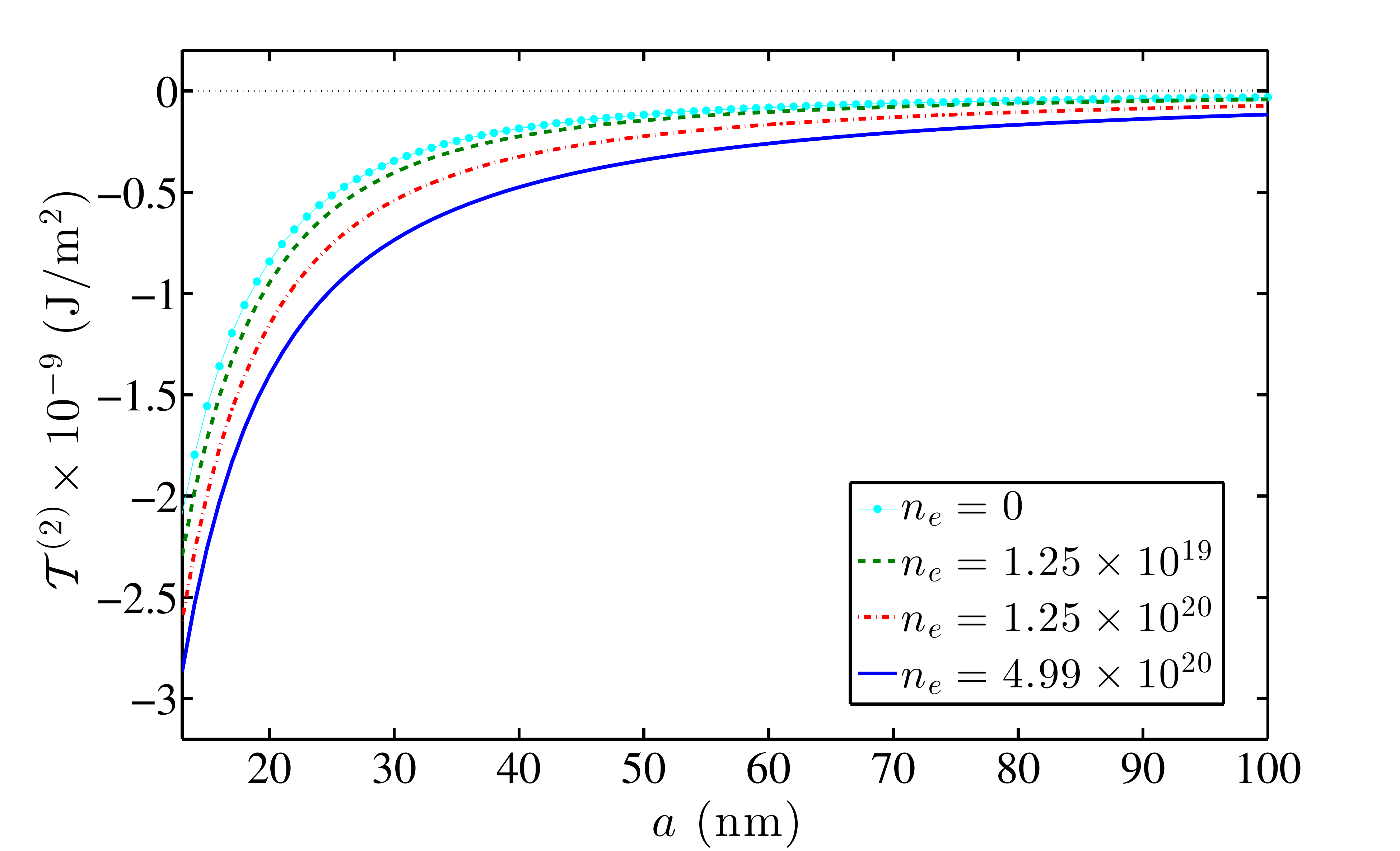}
\caption{(Color online) Torque per unit area as a function of separation distance between identical P slabs and identical carrier-excess slabs in vacuum ($XY$-$XY$ configuration). $n_e$ is in cm$^{-3}$.}
\label{torque_bulkci}
\end{center}
\end{figure}
%-------------
%--------------
\begin{figure}
\begin{center}
\includegraphics[width=9cm]{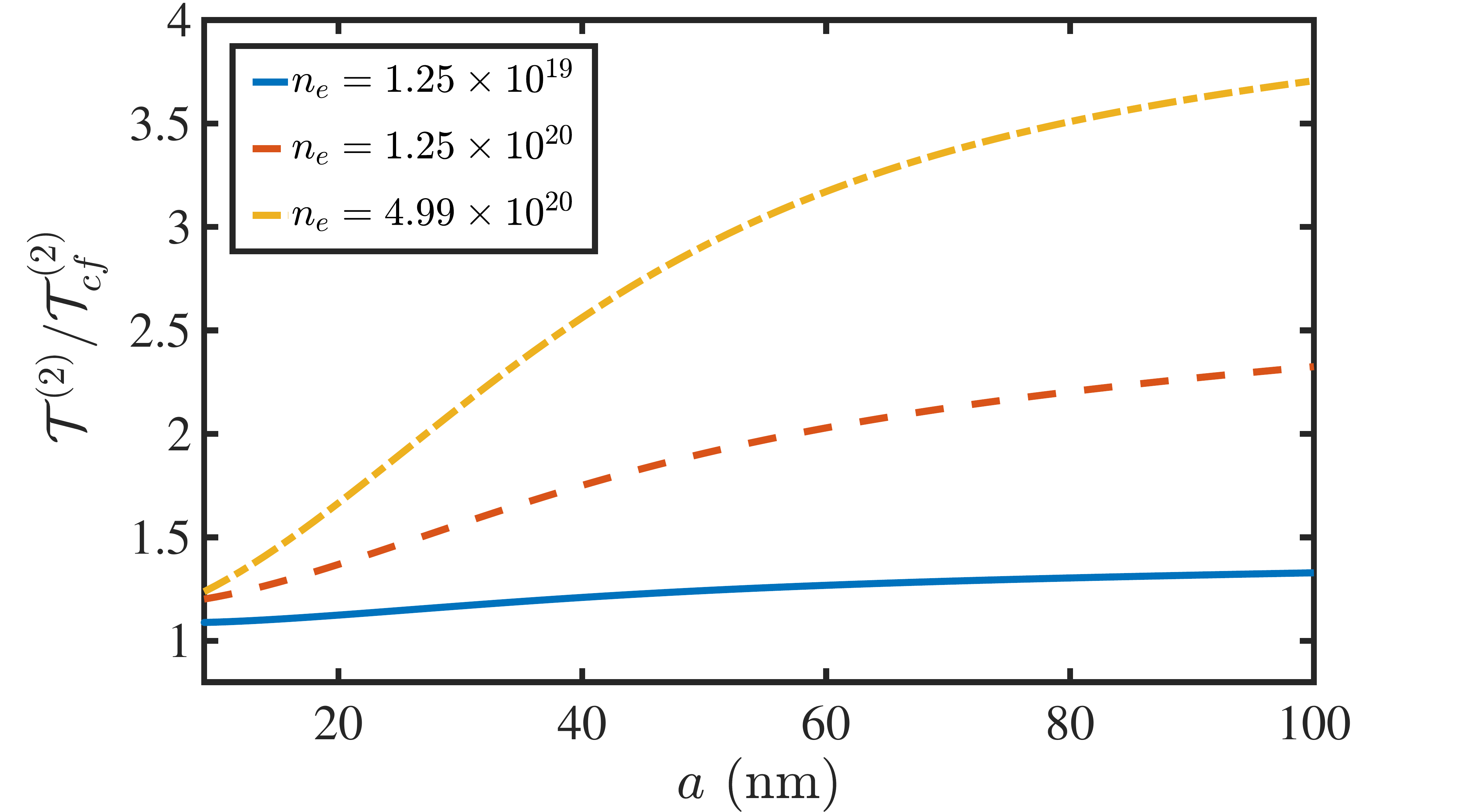}
\caption{(Color online) Ratio of the carrier-excess torque to the carrier-free torque as a function of the separation distance in vacuum ($XY$-$XY$ configuration). $n_e$ is in cm$^{-3}$.}
\label{torque_bulkci_ratio}
\end{center}
\end{figure}
%-------------

We first evaluate the torque between a pair of identical carrier-free black P slabs and between identical carrier-excess black P slabs for each carrier concentration at room temperature. The leading-order contribution to the torque, as given by Eq.~(\ref{tor-def}), for the four cases is plotted in Fig.~\ref{torque_bulkci}. It is observed that the higher the concentration of carriers, the larger is the magnitude of the torque. In Fig.~\ref{torque_bulkci_ratio}, we display the plots of the ratio of the carrier-excess torque and the carrier-free torque (${\mathcal{T}^{(2)}_{cf}}$). With extra carriers corresponding to $1.25\times$10$^{19}$ per cm$^{3}$, 
the magnitude of the torque is greater than that of the original identical carrier-free bulk slab interaction at all separations. 
The magnitude of the torque is even larger for the case of $n_e=1.25\times$10$^{20}$ cm$^{-3}$, which is twice as strong as the torque between a pair of original P slabs beyond $\sim$60 nm.
For $n_e=4.99\times$10$^{20}$ cm$^{-3}$, it becomes more than three times as large beyond $\sim$55 nm. 
That is, while the magnitude of torque itself decreases with distance, the degree of carrier enhancement increases with distance. This seems to be a consequence of the fact that carrier addition increases the dielectric values at lower Matsubara frequencies. 
The results indicate that an enhancement of the magnitude of the torque is attainable by means of carrier insertion.

\subsection{Non-identical configuration}

\begin{figure}
\begin{center}
\includegraphics[width=9cm]{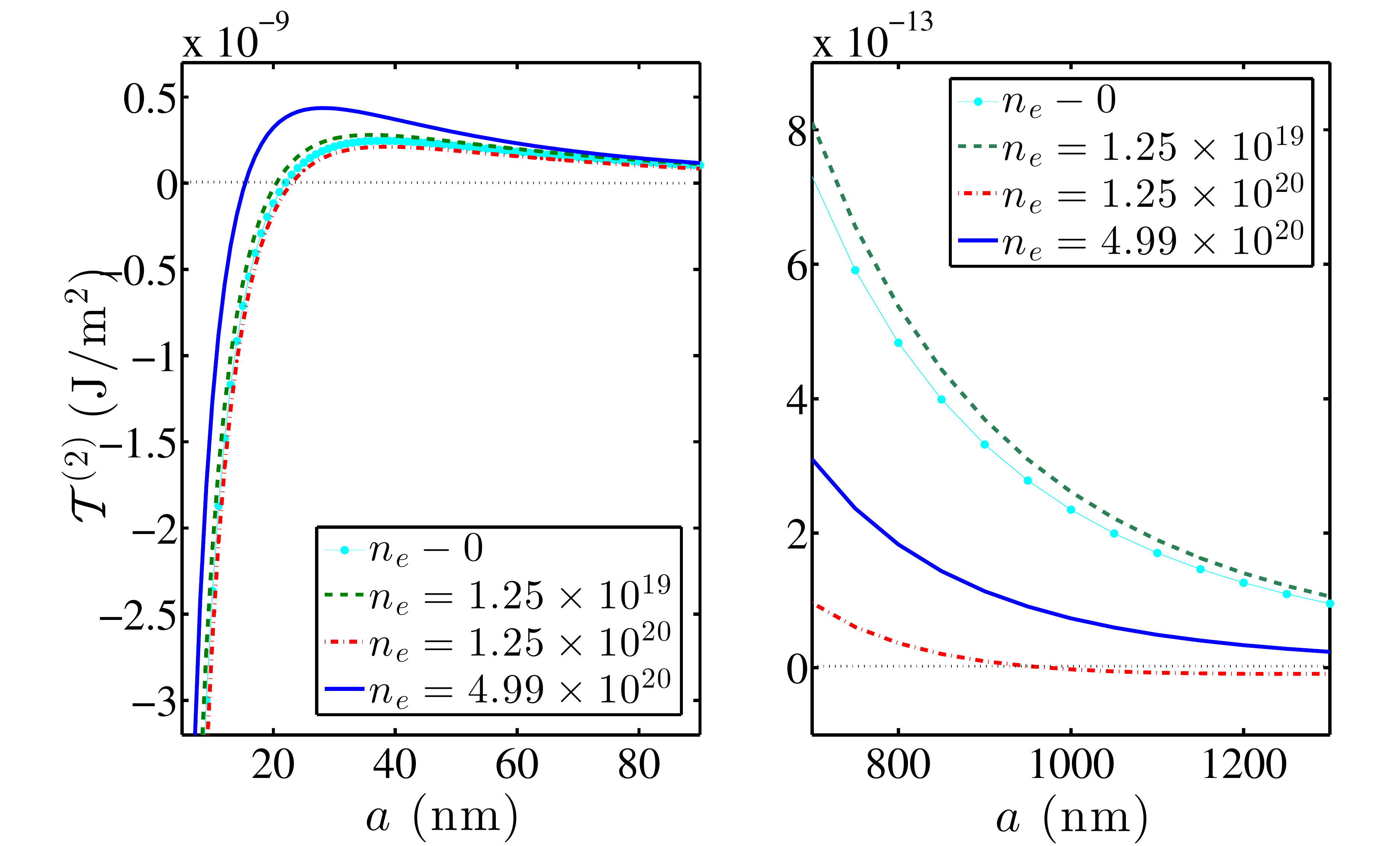}
\caption{(Color online) The torque per unit area as a function of distance when the $XY$ plane of the first slab interacts with the $YZ$ plane of the second slab, shown here for the carrier-free and carrier-excess P dielectric tensors. $n_e$ is in cm$^{-3}$. The two panels are for different distance regimes.}
\label{XY_YZ}
\end{center}
\end{figure}

Next, we consider the torque between non-identical configurations of the slabs. We have pointed out earlier that an intersection of the planar components of the dielectric tensor of a biaxial material when interacting with another biaxial material with non-crossing planar components can render a reversal of the sign of torque between them~\cite{PRL2018}. The sign reversal is possible when the second material also has a crossing of the planar components but at a frequency significantly different from that of the crossing of the first material. Below and above this critical frequency at which $\beta=0$, the contributions to the torque will be of opposite signs. The overall sign of the torque at a fixed separation distance is then determined by the summation of the contributions to torque from each frequency. 
We will first consider the configuration when the $XY$ plane of one slab interacts with the $YZ$ plane of the other slab. That is, the face $XY$ of the first slab and the face $YZ$ of the second slab are set parallel to each other and perpendicular to the normal direction $\hat {\bf n}$. In this configuration, the carrier-inserted dielectric tensor with $n_e=1.25\times$10$^{20}$ cm$^{-3}$ has a crossing of the planar components on each face at frequencies $\zeta_{xy}$ and $\zeta_{yz}$ respectively. We observe a reversal of the sign of torque twice with separation distance corresponding to each intersection as shown in Fig.~\ref{XY_YZ}. For $n_e=1.25\times 10^{19}$ cm$^{-3}$ and $4.99\times 10^{20}$   cm$^{-3}$, the sign of torque changes once corresponding to the crossing $\zeta_{xy}$ on the $XY$ face. 

%--------------
%-------------

%--------------
\begin{figure}
\begin{center}
\includegraphics[width=8.9cm]{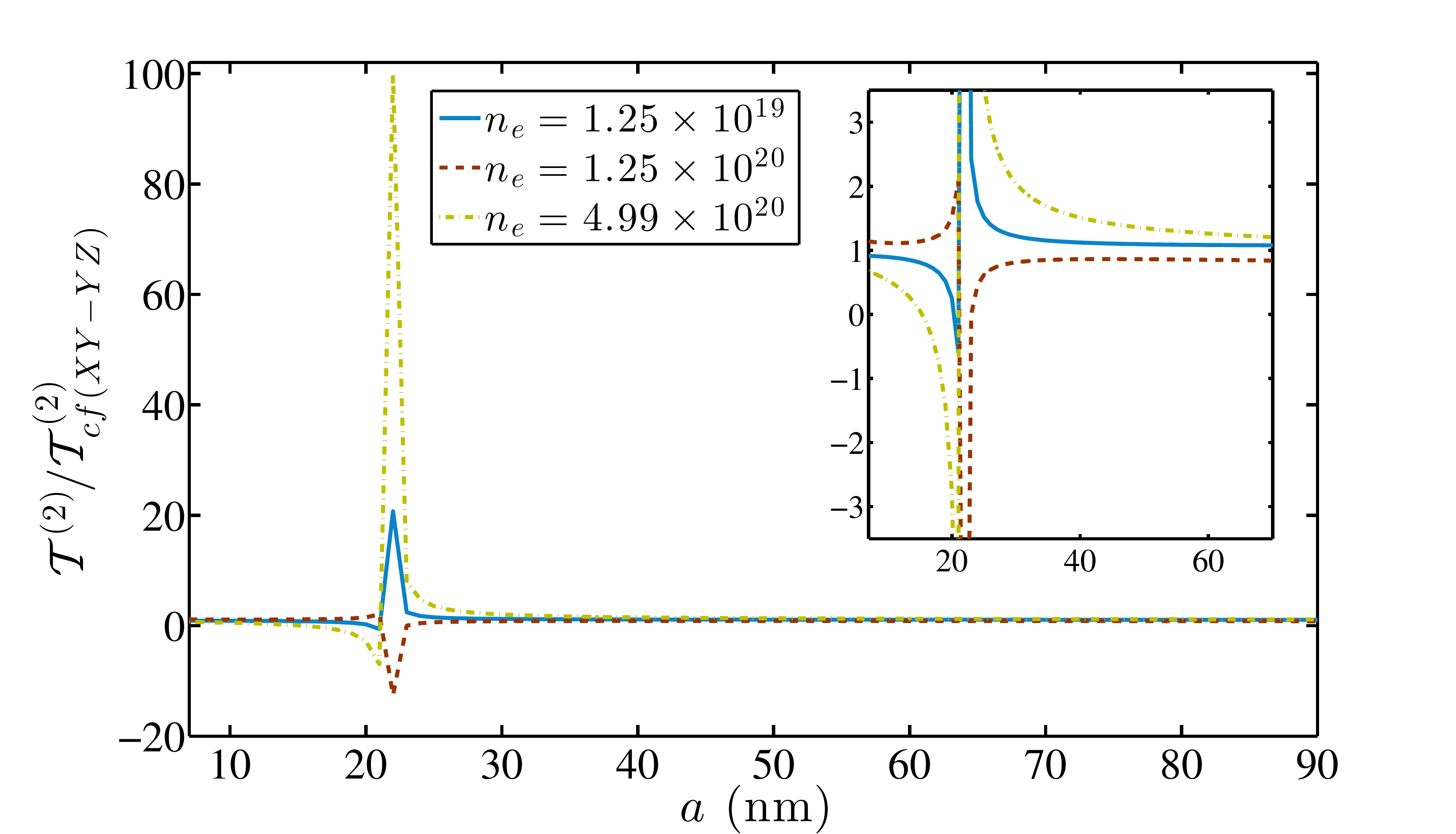}
\caption{(Color online) Ratio of the carrier-excess torque to the carrier-free torque as a function of the separation distance in the $XY$-$YZ$ configuration in vacuum. $n_e$ is in cm$^{-3}$.}
\label{ratioXYYZ}
\end{center}
\end{figure}
%-------------

The magnitudes of torque in this configuration are quite different compared to the case of identical slab interaction. At smaller separations, the carrier-excess slabs with greater concentrations of extra carriers start out with smaller magnitudes of torque but they change sign sooner and exhibit higher peaks on the positive side of the axis; see Fig.~\ref{XY_YZ}. An exception is the case of $n_e=1.25\times$10$^{20}$ cm$^{-3}$ displaying the largest magnitude of torque on the negative side and the least magnitude on the positive side that gradually goes to zero again to exhibit the second sign reversal. The magnitude of the torque thus sensitively depends on the cancellation between the positive and the negative contributions arising from the in-planar crossing of the dielectric tensors. The plot of the ratios with respect to the carrier-free torque in this orientation, ${\mathcal{T}^{(2)}_{cf (XY-YZ)}}$ is depicted in Fig.~\ref{ratioXYYZ}. Due to the sign-reversal effect, the increase in the degree of enhancement with distance that we earlier observed for the identical orientation (see Fig.~\ref{torque_bulkci_ratio}) is no longer seen here. The peaks in this figure do not have any significance regarding relative magnitudes of the torque, and correspond only to the distance where the carrier-free torque, ${\mathcal{T}^{(2)}_{cf (XY-YZ)}}$ tends to zero and changes sign.

\begin{figure}
\begin{center}
\includegraphics[width=9cm]{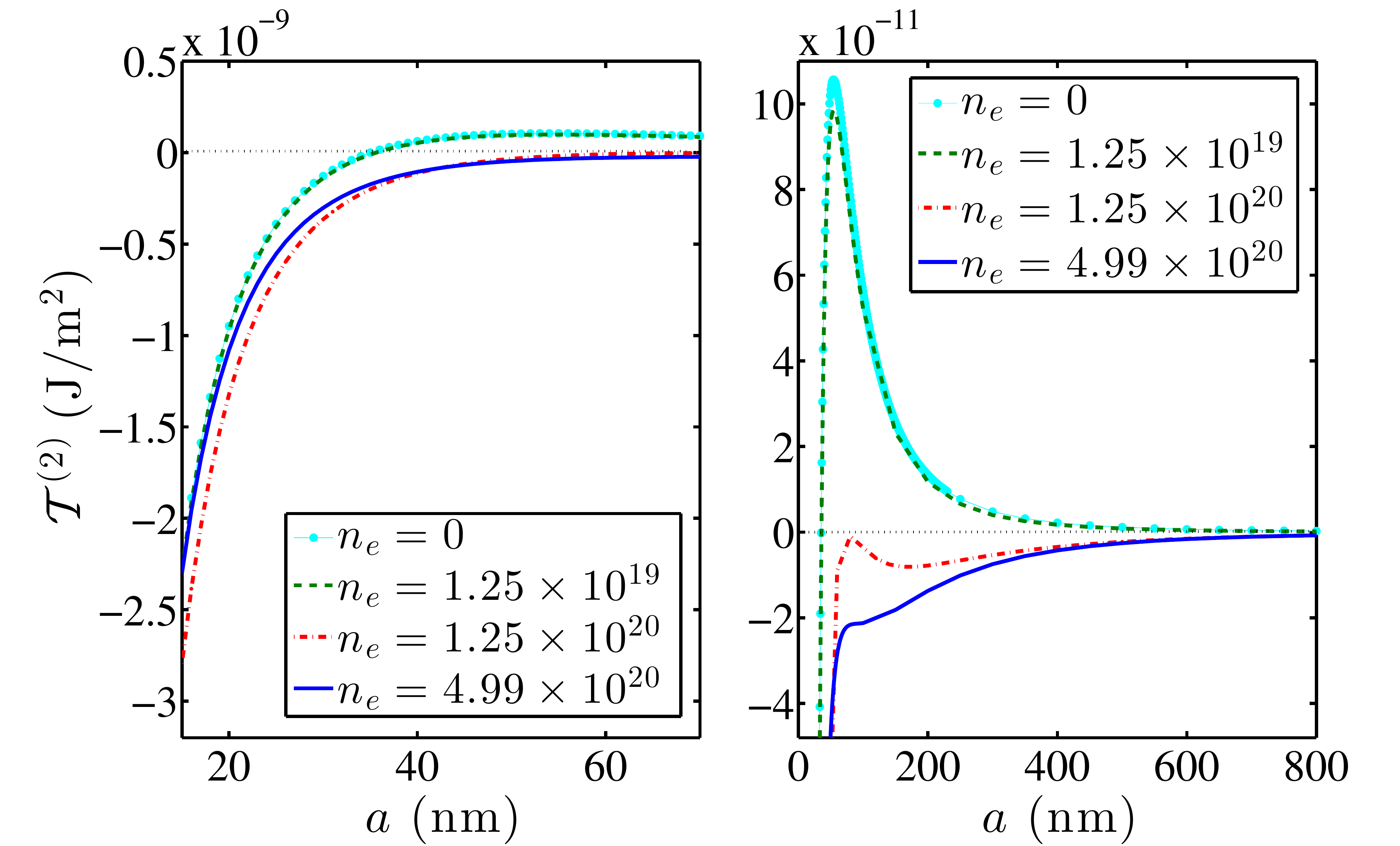}
\caption{(Color online) Torque per unit area as a function of distance when the $XY$ plane of the first slab interacts with the $XZ$ plane of the second slab, shown here for the carrier-free and carrier-excess P dielectric tensors. $n_e$ is in cm$^{-3}$.}
\label{XY_XZ}
\end{center}
\end{figure}

In Fig.~\ref{XY_XZ}, we show the torque between the $XY$ and $XZ$ planes. In this configuration also, each face has an intersection of in-planar dielectric components at $\zeta_{xy}$ and $\zeta_{xz}$ respectively for the carrier-excess dielectric tensors with $n_e=1.25\times 10^{20}$ cm$^{-3}$ and $4.99\times 10^{20}$ cm$^{-3}$. However, we observe no double reversal of sign. Moreover, the change of sign corresponding to the $\zeta_{xy}$ crossing seen for $n_e=0$ and $1.25\times 10^{19}$ cm$^{-3}$ is suppressed in the case of $n_e=1.25\times 10^{20}$ cm$^{-3}$ and $4.99\times 10^{20}$ cm$^{-3}$. As displayed in Fig.~\ref{XY_XZ}, the curves for $n_e=1.25\times 10^{20}$ cm$^{-3}$ and $4.99\times 10^{20}$ cm$^{-3}$ do show non-monotonic features with extrema, but the negative contributions to the torque dominate and curb the sign reversal. This emphasizes that the property of in-planar crossing of dielectric components is a necessary but not sufficient prerequisite for the prediction of sign reversal. 
A detailed computation of the contributions from each Matsubara frequency mode is always required. This also explains why the sign-reversal with distance did not show up in the experimental measurement of the torque between calcite (CaCO$_3$) and a liquid crystal (5CB) in Ref.~\cite{SomersNature2018}. The crossing of the in-planar frequencies of CaCO$_3$ occurs too early, and the contribution to the torque before the crossing is not sufficient to dominate the contribution to the torque after the crossing. Hence, the overall torque will bear the sign of the dominating side throughout and will not change sign with distance. For a suitable choice of a pair of materials, it is our expectation that the sign-reversal will appear in experiments.

%--------------
%-------------

%--------------
\begin{figure}
\begin{center}
\includegraphics[width=9cm]{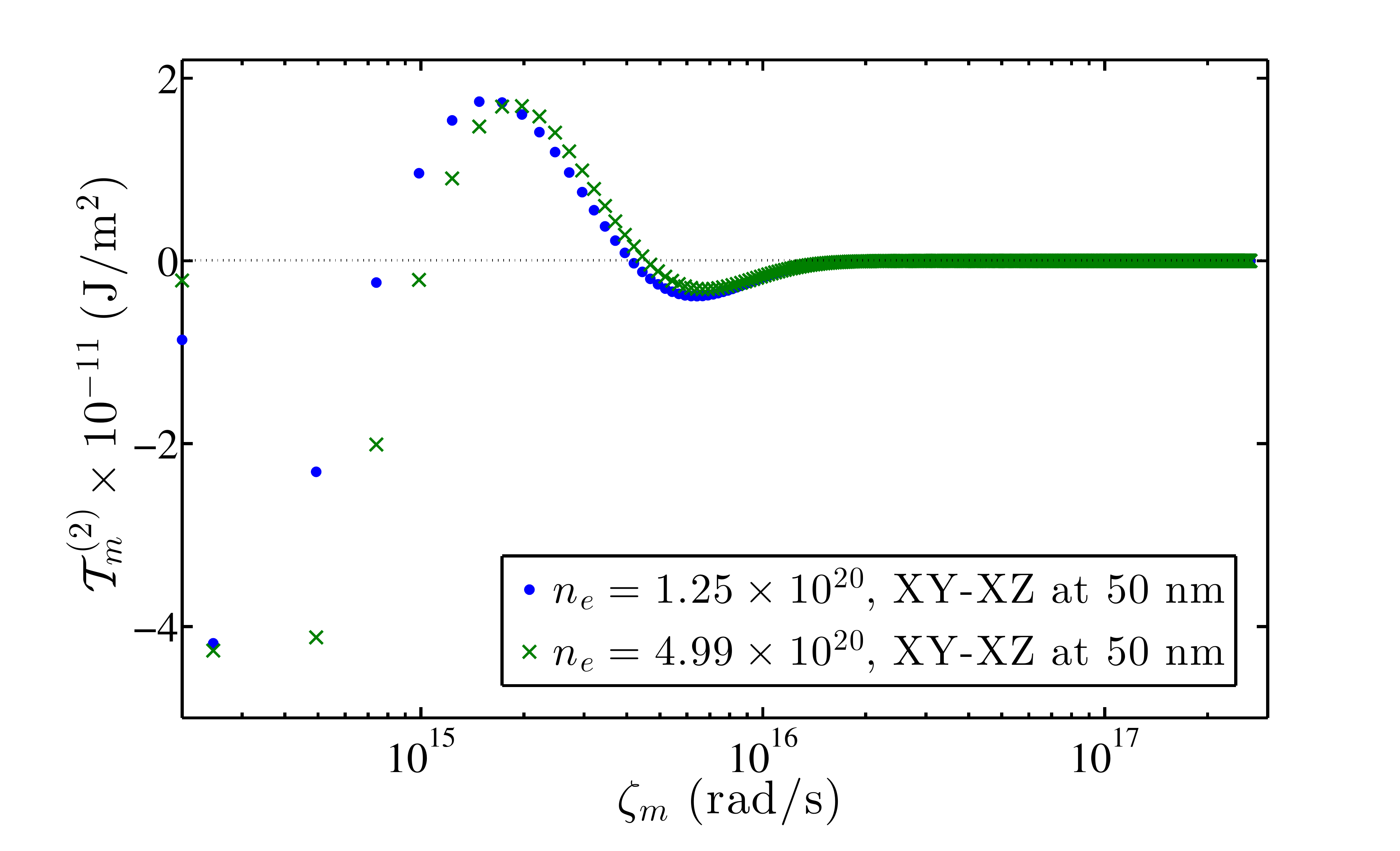}
\caption{(Color online) Spectral plot showing the contribution to the torque from each Matsubara frequency $\mathcal{T}^{(2)}_m$ for the interaction between the $XY$ plane and $XZ$ plane fixed at a separation distance of 50 nm. The values at $m=0$ are placed on the $y$-axis. $n_e$ is in cm$^{-3}$.}
\label{spectral_fn}
\end{center}
\end{figure}
%-------------

\begin{figure}
\centering
\includegraphics[width=8.5cm]{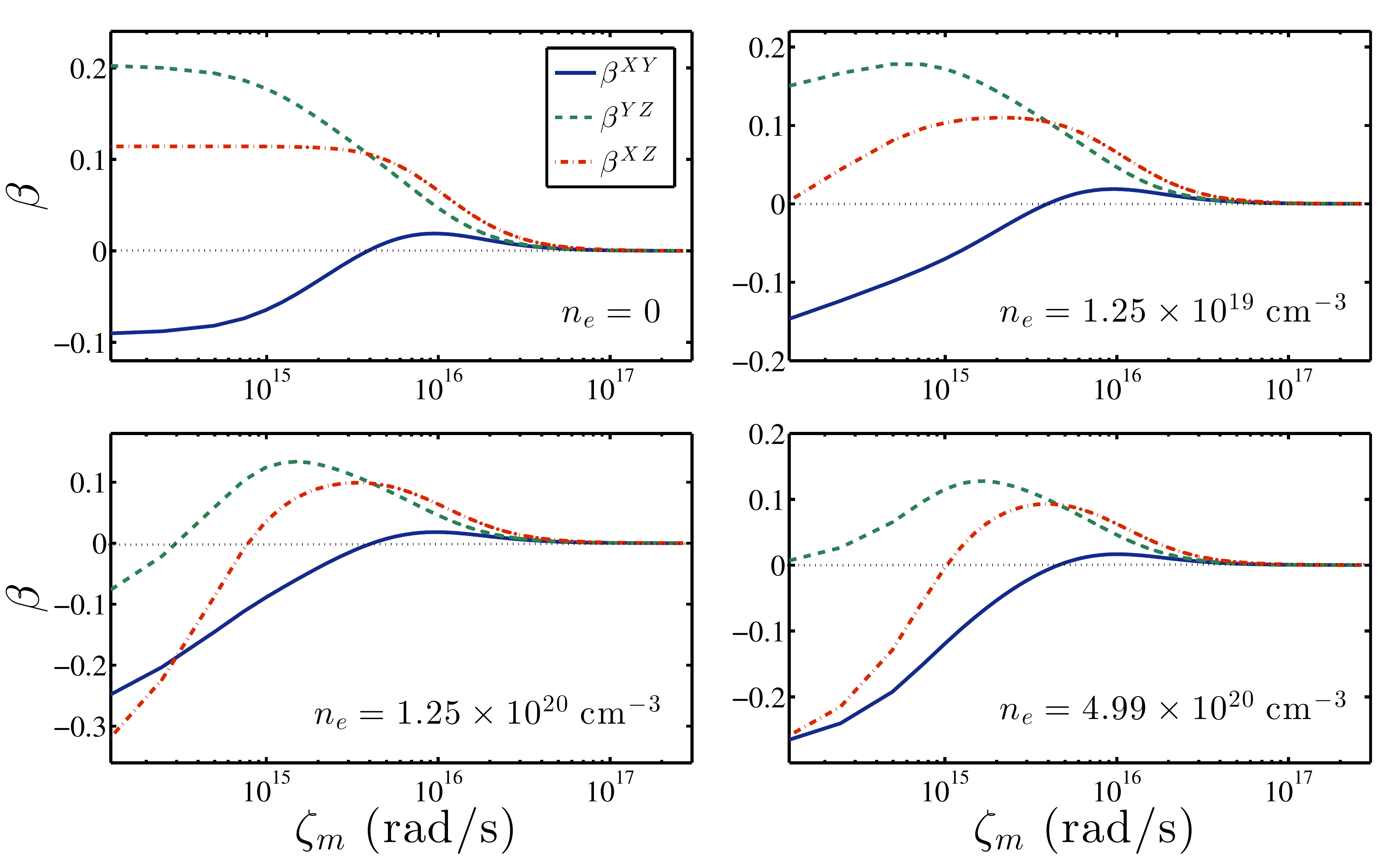}
\caption{(Color online) Plot of the perturbative parameter $\beta$ as given by Eq.~(\ref{beta_para}) for the $XY$, $YZ$ and $XZ$ planes of the carrier-free bulk P dielectric tensor and the different carrier-excess dielectric tensors. The values at $m=0$ are placed on the $y$-axis.}
\label{beta}
\end{figure}

The spectral plot in Fig.~\ref{spectral_fn} of each frequency mode contribution $\mathcal{T}^{(2)}_m$ illustrates why we observe no sign reversal in the expected cases of $n_e=1.25\times 10^{20}$ cm$^{-3}$ and $4.99\times 10^{20}$ cm$^{-3}$.
The figure indicates that the contribution from each Matsubara frequency is important. At a separation of 50 nm, for instance, the area under the positive curves remains smaller than the area under the negative curves, and the torque stays negative. 
In this regard, the plot of the perturbative parameter $\beta$ as given by Eq.~(\ref{beta_para}) may throw a little insight on the expectation of sign reversal. $\beta$ for the different faces of the original and the carrier-excess P is displayed in Fig.~\ref{beta}. Where there is no crossing of the planar dielectric components on either of the two interacting faces, the product of the $\beta$ parameters for the two faces retains the same sign throughout the whole frequency regime, and a sign reversal is not predicted. Where there is a crossing of the planar components, the product of the $\beta$ parameters possesses opposite signs above and below the critical crossing frequency. The contribution from the $k$-integral is always positive.
The sign of $\beta_1\beta_2$ coupled with the contributions from the $k$-integral, and summed over all frequencies will finally determine the overall sign of torque, as given by Eq.~\ref{tor-def}.

%\begin{widetext}
\hspace{7px}
%\end{widetext}

%
%

\begin{figure}
\begin{center}
\includegraphics[width=9cm]{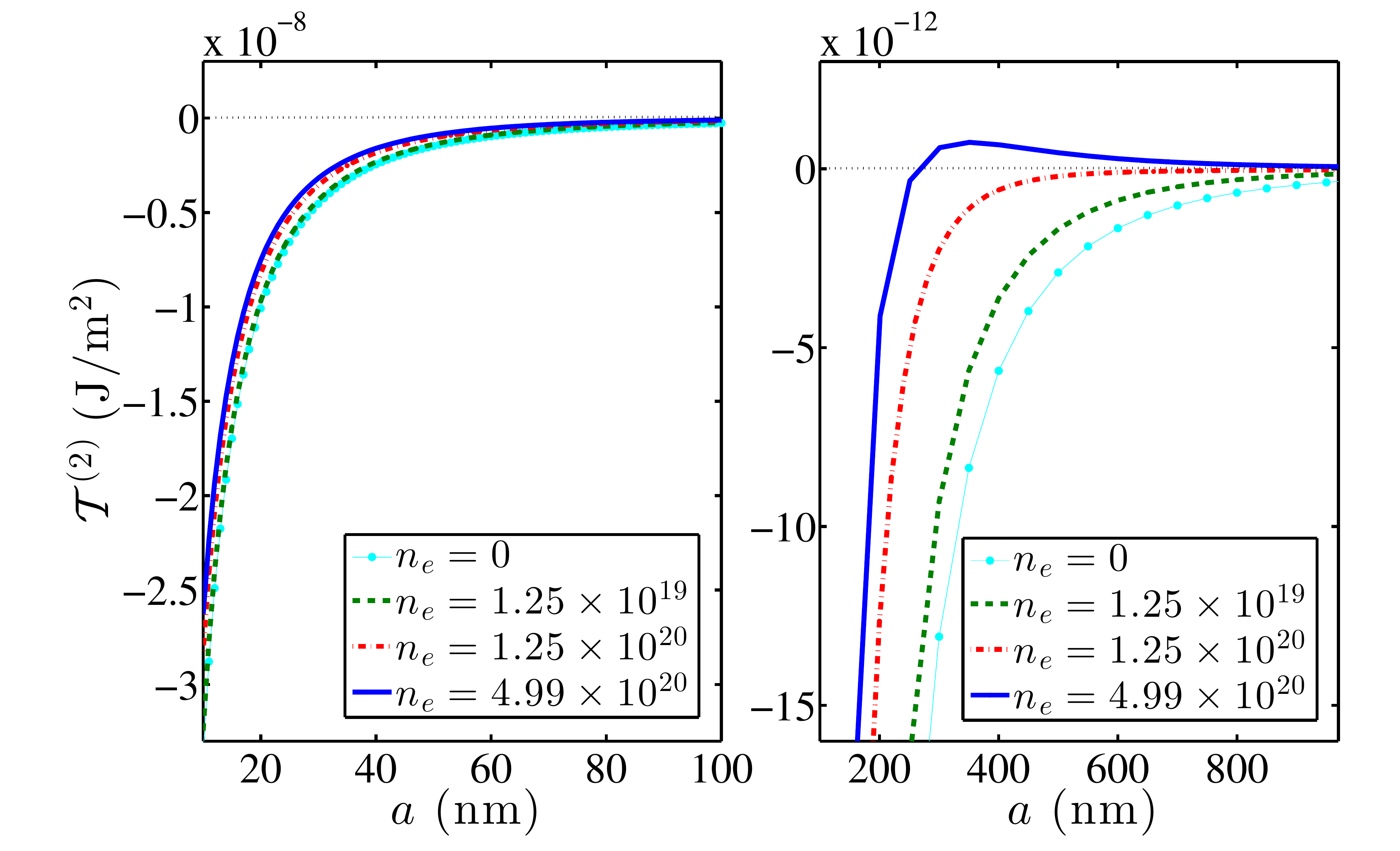}
\caption{(Color online) Torque per unit area as a function of distance when the $YZ$ plane of the first slab interacts with the $XZ$ plane of the second slab, shown here for the carrier-free and carrier-excess P dielectric tensors.}
\label{YZ_XZ}
\end{center}
\end{figure}

A plot of the torque for the remaining configuration when the $YZ$ plane faces the $XZ$ plane is shown in Fig.~\ref{YZ_XZ}. In this case, the sign reversal occurs for $n_e=4.99\times 10^{20}$ cm$^{-3}$, but not for the other investigated concentrations. The striking feature of this configuration lies in the magnitude of the torque which is almost an order larger than for the rest of the configurations. 

\begin{figure}
\begin{center}
\includegraphics[width=9cm]{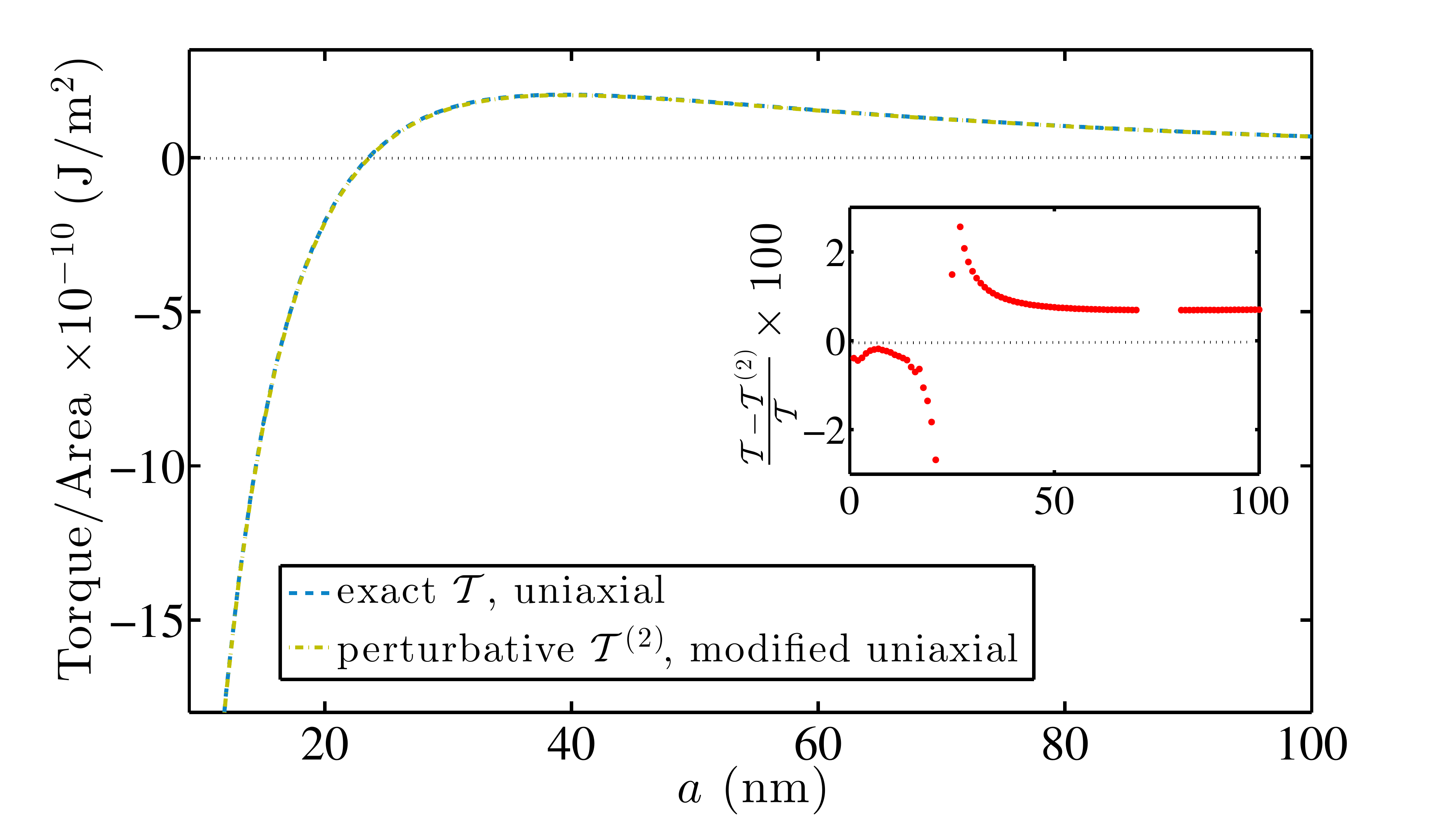}
\caption{(Color online) Comparison of the leading order torque in the uniaxial limit (setting $\varepsilon^n=\varepsilon^v$) and Barash's exact result in the $XY$-$YZ$ orientation of black P with $n_e=1.25\times$10$^{20}$ cm$^{-3}$. The inset shows the relative percentage error. }
\label{uni_compare}
\end{center}
\end{figure}

The quantitative results presented here are, however, subject to limitations of the modeling of carrier injection by DFT and the subsequent dielectric tensors. It is also subject to how well the perturbative theory estimates the torque in the leading order. In this regard, we compare the leading order torque in the uniaxial limit with the exact uniaxial result of Barash's shown for one case ($XY$-$YZ$) in Fig.~\ref{uni_compare}. For this purpose, we set the dielectric component in the normal direction $\hat{\bf n}$ equal to one of the in-planar dielectric components $\varepsilon^n=\varepsilon^v$ 
%that is $\varepsilon^z=\varepsilon^y$
in black P. Figure~\ref{uni_compare} illustrates the excellent agreement between the two theories for black P with the relative error remaining mostly below 2 {\%}. Since we consider the $XY$-$YZ$ orientation, the relative error tends to grow near the separation distance where the torque changes sign, which is expected as the denominator tends to zero. The agreement turns out to be equally good for the other orientations as well. Thus, the leading order contribution to the torque is dominant, and results in faster computation than the exact calculation. For example, in a laptop using Mathematica, the perturbative calculation runs more than 50 times faster than the corresponding exact calculation.

%--------------
%-------------

%--------------
%-------------

{\sout {Theoretically,}} The distance at which the sign changes could be roughly estimated as $a_{c, {\mathrm{est}}}=c/2\zeta_c$, where $\zeta_c=\zeta_{xy}$ or $\zeta_{yz}$ or $\zeta_{xz}$~\cite{PRB2017}. {\tcc {The theoretical distance $a_{c, {\mathrm{th}}}$ obtained from our perturbative calculations and the estimated distance $a_{c, {\mathrm{est}}}$}} at which sign-change occurs are compared in Table~\ref{crossing_dist}. This is a simple order of magnitude estimation based on the crossing frequency of the planar dielectric components, and as seen from Table~\ref{crossing_dist}, it works qualitatively for the anisotropically polarizable materials as well.
Note again from this table that the sign-reversal does not always show up where it is expected.

\begin{table}
\begin{center}
\caption{The estimated distance $a_{c,{\mathrm{est}}}$ (nm) and the theoretical distance $a_{c,{\mathrm{th}}}$ (nm) at which the torque reverses sign corresponding to each crossing frequency in the three configurations of interaction between different faces of black P.}
%\centering
\small\addtolength{\tabcolsep}{3.5pt}
\label{crossing_dist}
\vspace{5px}
%\begin{adjustbox}{max width=\textwidth}
\begin{tabular}{ccccccc}
\hline
\hline
\multicolumn{1}{c}{} &\multicolumn{2}{c}{$XY$-$YZ$} &\multicolumn{2}{c}{$XY$-$XZ$} &\multicolumn{2}{c}{$YZ$-$XZ$}\\
\cline{2-7}
$n_e$ (cm$^{-3}$)& \multicolumn{1}{c}{$a_{c, {\mathrm{est}}}$} & \multicolumn{1}{c}{$a_{c, {\mathrm{th}}}$} &\multicolumn{1}{c}{$a_{c, {\mathrm{est}}}$} & \multicolumn{1}{c}{$a_{c, {\mathrm{th}}}$}& \multicolumn{1}{c}{$a_{c, {\mathrm{est}}}$} & \multicolumn{1}{c}{$a_{c, {\mathrm{th}}}$}\\
\hline
0      & 38.5  &22  & 38.5 &36 &- &-   \\
              
\hline
$1.25 \times$10$^{19}$      & 38.1   & 21  & 38.1  & 36 &- &-   \\
  
 \hline
$1.25 \times$10$^{20}$      &36.3   & 23 & 36.3 &- & 455.9 & -   \\
             & 455.9  & 1000  & 189.9  &- &189.9 &-   \\

 \hline
$4.99 \times$10$^{20}$      &33.3   &16  &33.3  &- &146 & 301  \\
            &-   &-  &146  &- &- &-   \\

\hline
\hline
\end{tabular}
%\end{adjustbox}
\end{center}
\end{table}

\section{Effect of intervening medium}

\begin{figure}
\begin{center}
\includegraphics[width=9cm]{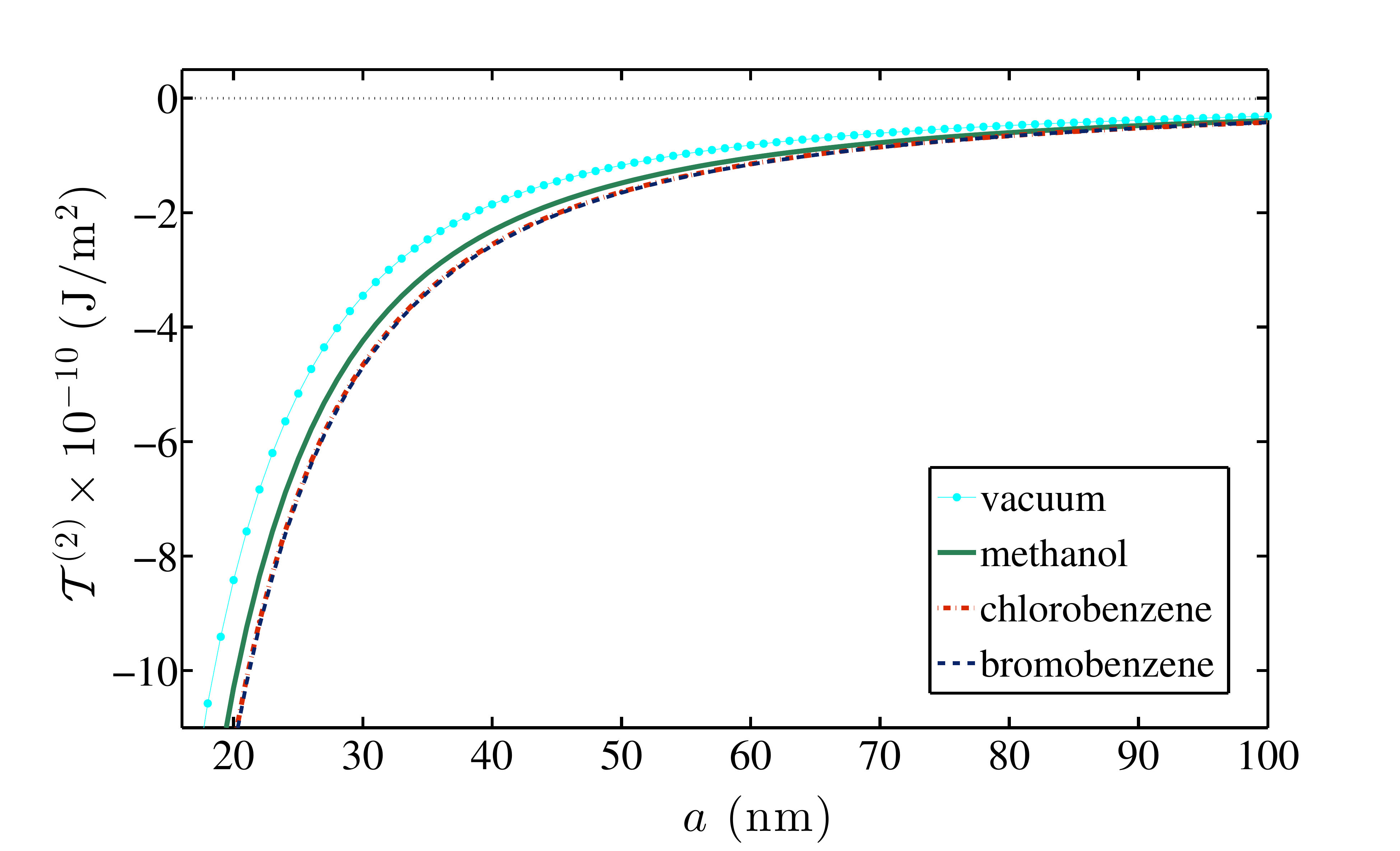}
\caption{(Color online) Torque per unit area as a function of separation distance between original identical P slabs ($XY$-$XY$ orientation, $n_e=0$) in various intervening isotropic media.}
\label{torque_medium}
\end{center}
\end{figure}

\begin{figure}
\begin{center}
\includegraphics[width=9cm]{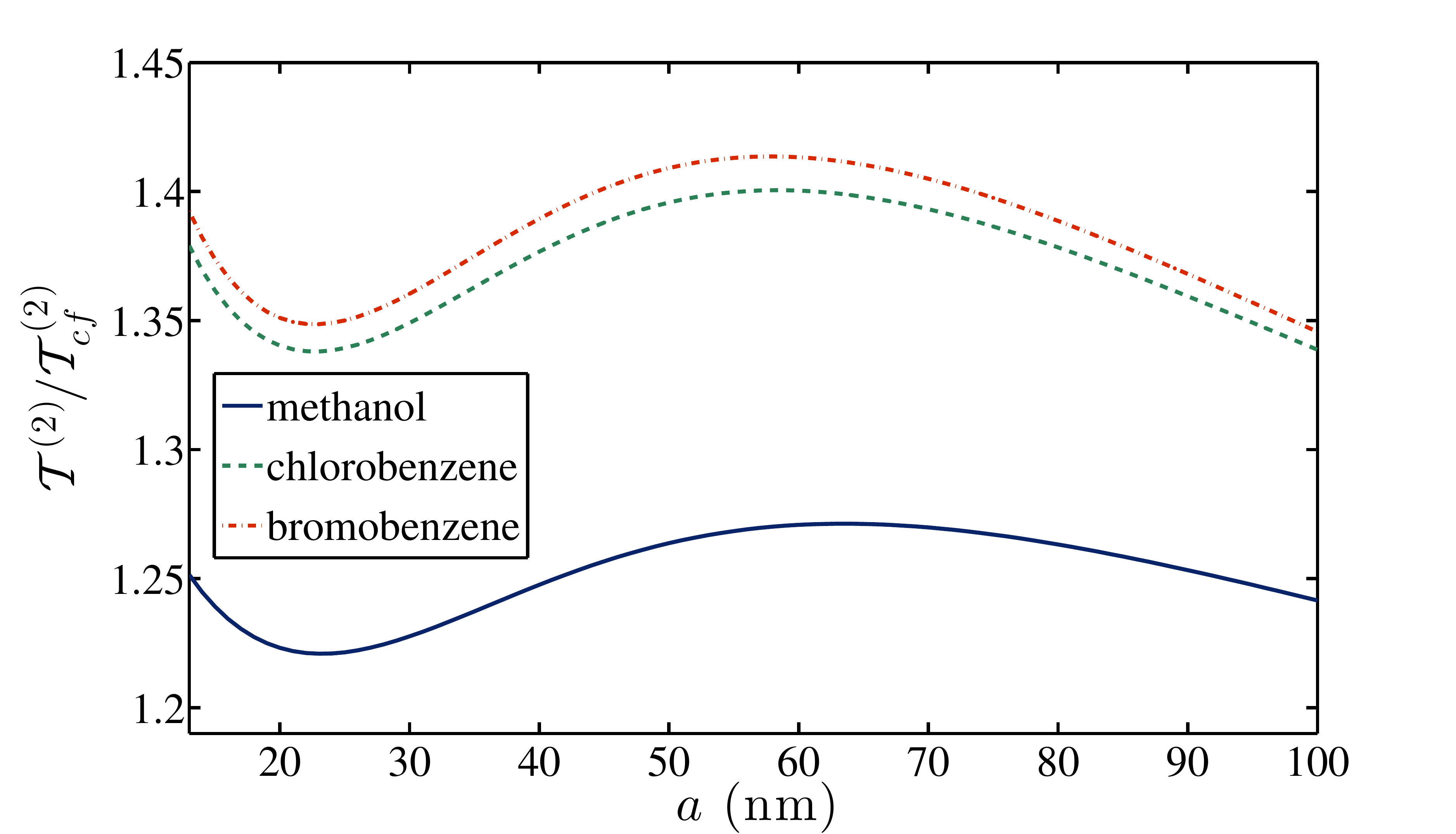}
\caption{(Color online) Ratio of the carrier-free torque in medium to the carrier-free torque in vacuum as a function of the separation distance for the $XY$-$XY$ orientated slabs.}
\label{torque_medium_ratio}
\end{center}
\end{figure}

The effect of an intervening isotropic medium on the torque between a pair of interacting slabs is also considered in this study. Usually, an intervening medium reduces the normal attractive interaction between materials. However, the torque between two anisotropic materials is shown here to increase, as was also observed in Ref.~\cite{SomersPRL2018}. For the purpose of illustration, we consider methanol, chlorobenzene and bromobenzene as the background intervening media. 
The dielectric functions of the media under consideration here are obtained from Ref.~\cite{ZwolPalasantzas}. Figure~\ref{torque_medium} compares the torque per unit area between identical carrier-free P slabs in vacuum and in medium. Enhancement is observed for all the media and at all separation distances. The strength of enhancement, however, does not reach a factor of 2.
The plot of the ratios of the torque in the media with respect to that in vacuum is shown in Fig.~\ref{torque_medium_ratio}. An interesting result noted here is the non-monotonous degree of enhancement with respect to distance even in the identical configuration of the interacting slabs without any excess charge carriers. {\tcc {An intervening liquid will give rise to capillary adhesion and hydrodynamic forces. These are under usual circumstances directed normal to the plates, and  will  thus   not interfere with the calculated torques. High-index intervening liquids have been   studied  extensively in a repulsive Casimir context~\cite{ZwolPalasantzas}.}}

%--------------
%-------------

%--------------
%-------------

\begin{figure}
\begin{center}
\includegraphics[width=9.5cm]{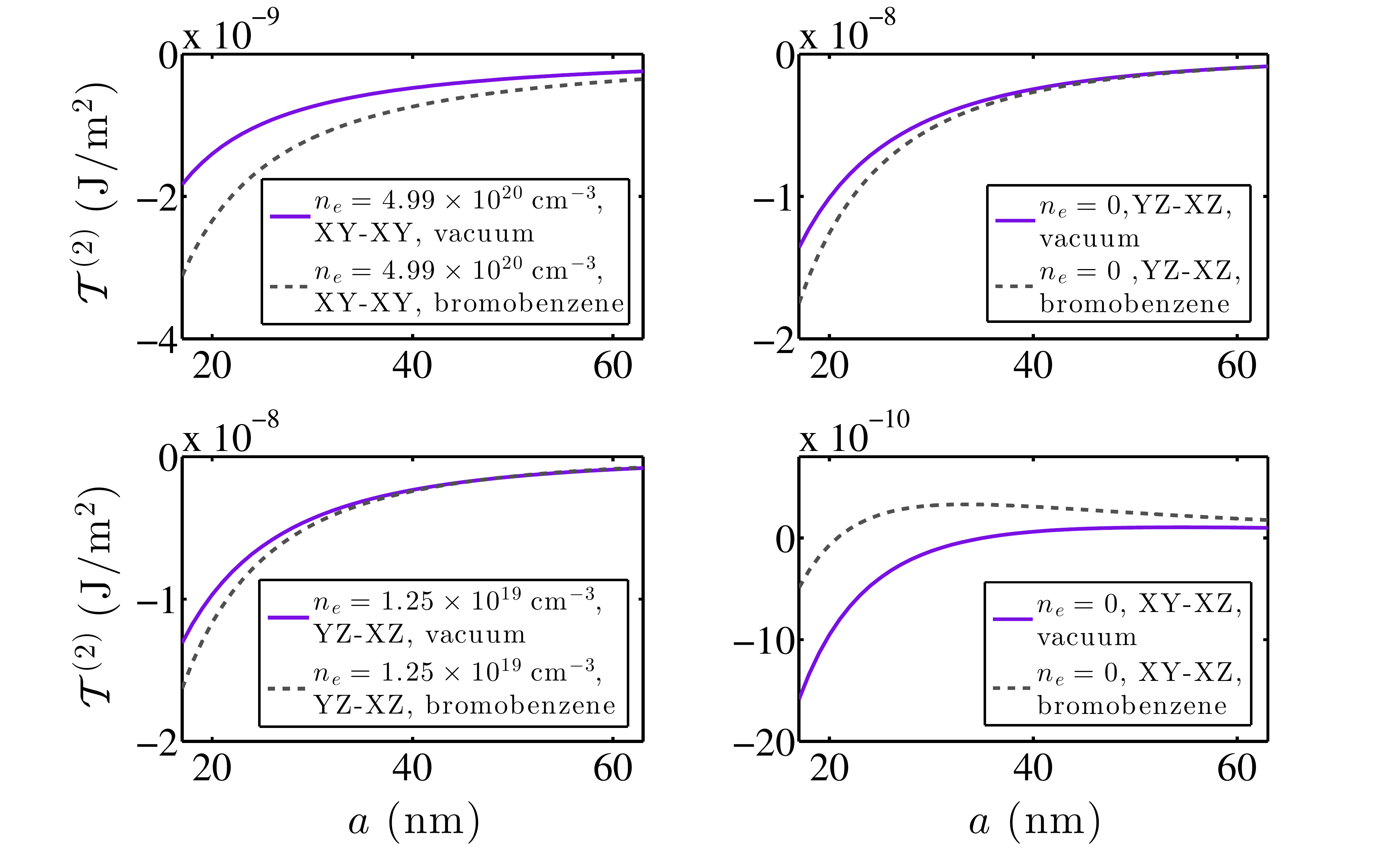}
\caption{(Color online) Comparison of the torque at different configurations in the presence and absence of the medium. }
\label{torque_large}
\end{center}
\end{figure}

In Fig.~\ref{torque_large}, we compare magnitudes at different orientations of the carrier-free as well as carrier-excess P slabs in the presence and the absence of the medium. A medium induces enhancement in most cases. However, at certain orientations such as the $XY$-$XZ$, the medium instead acts to induce faster reversal of the sign of the torque without enhancement. It is interesting that the original bulk P slabs at the orientation $YZ$-$XZ$ without any extra carriers turns out to be most effective for torque enhancement when a medium is involved. At the same orientation, addition of extra carriers of $n_e=1.25\times10^{-19}$ cm$^{-3}$ to black P results in slight reduction of the torque magnitude in bromobenzene; compare bottom left and top right panels of Fig.~\ref{torque_large}.
This may in fact be a favourable feature in view of experimental tests, since involvement of both carrier injection and intervening medium may lead to complications. {\tcc {However, this is not a general result as we see below}}.

%--------------
%-------------

\section{Enhancement}

\begin{figure}
\begin{center}
\includegraphics[width=9cm]{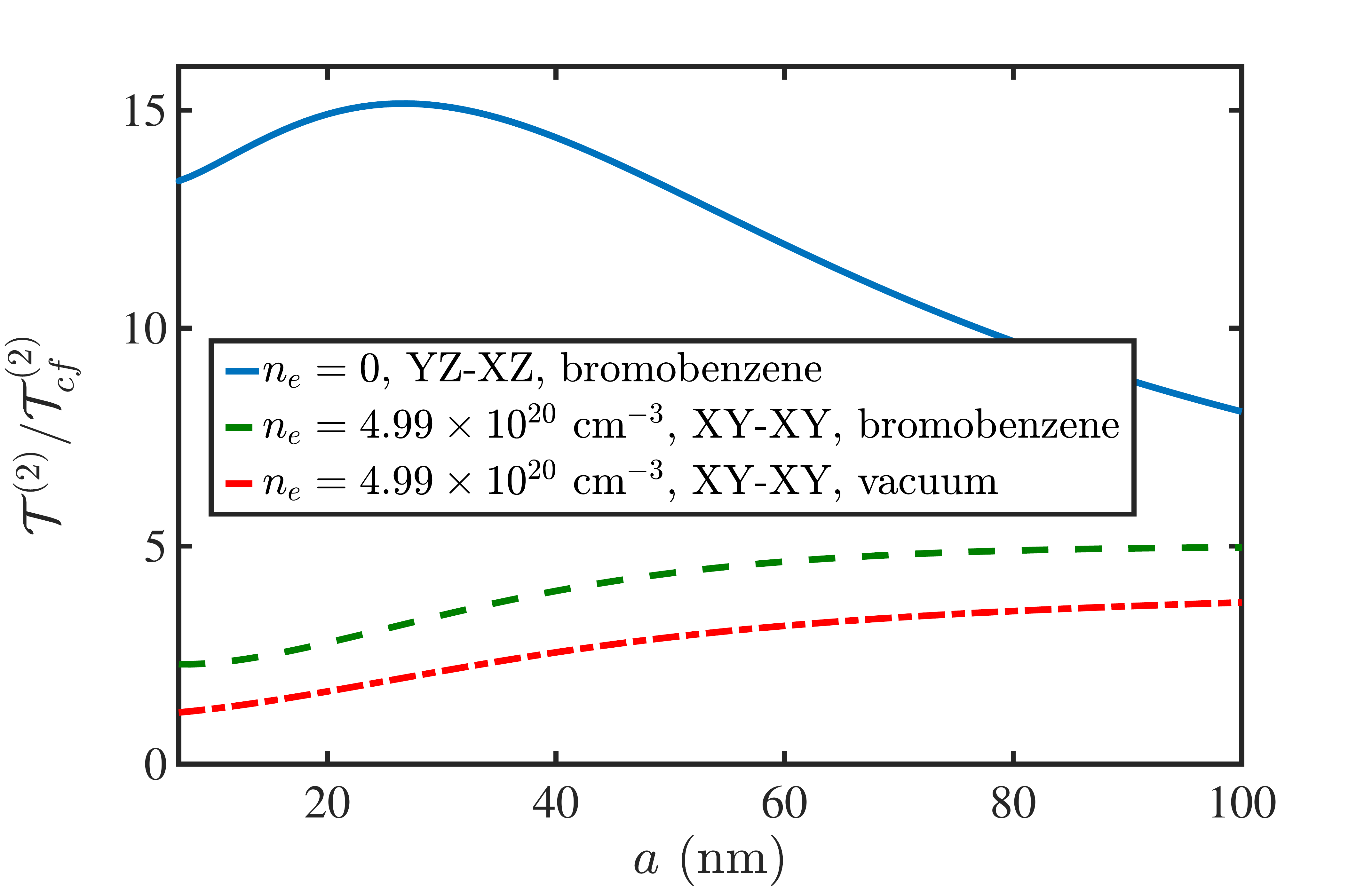}
\caption{(Color online) A collection of cases with relatively larger magnitudes of torque. They involve a combination of carrier-excess dielectric functions, different configurations and intervening medium. {\tcc {The different configurations are compared with the identical ($XY$-$XY$), carrier-free P slabs in vacuum}}.  }
\label{percent_increase}
\end{center}
\end{figure}

Finally, the cases with the greatest degrees of enhancement are collected in this section and compared with the identical vacuum case. The ratios of the torque for these select cases involving both medium and excess charge carriers to the torque between identical carrier-free P slabs ($XY$-$XY$) in vacuum, $\mathcal{T}^{(2)}_{cf}$, are shown in Fig.~\ref{percent_increase} as a function of separation distance. At the $YZ$-$XZ$ orientation of the planes of the interacting P slabs in the background medium of bromobenzene, we observe occurrence of more than an order of magnitude enhancement of torque as compared with the identical ($XY$-$XY$) vacuum case in the absence of extra carriers; see the blue solid curve in Fig.~\ref{percent_increase}. As stated earlier, addition of extra carriers raises the degree of increase of torque with separation in vacuum in the identical $XY$-$XY$ orientation; see the red curve (dash-dotted). The increase in torque is further enhanced in the presence of bromobenzene as an intervening medium; see the green dashed curve. {\tcc {This is in contrast to what we have observed in Fig.~\ref{torque_large} for the $YZ$-$XZ$ configuration}}. The non-monotonic degree of enhancement that we earlier observed in Fig.~\ref{torque_medium_ratio} in the identical orientation in the presence of a medium no longer prevails, and the effect of excess charge carriers dominates. With the change in orientation of the slab shown here for $YZ$-$XZ$, non-monotonicity appears in the ratio with distance. This is caused by the tendency in this orientation towards sign reversal. Figure~\ref{percent_increase} highlights the possibilities of torque enhancement with suitable choices of interacting materials and intervening media.

%--------------
%-------------

\section{Conclusions}

In this work, we present an investigation on the sign-reversal behaviour of the Casimir-Lifshitz torque with separation distance using biaxially polarizable black P whose frequency-dependent dielectric tensor components are enhanced by introducing extra charge carriers. It is observed that the reversal of the sign of torque with separation distance does not always occur when the prerequisite of a crossing of the in-planar dielectric components of one of the interacting slabs is satisfied. The calculation needs to be performed to ascertain the prediction of sign-reversal. However, a sign-reversal with distance can take place only when the said prerequisite is satisfied. It is our expectation that, for a suitable choice of a pair of dielectric materials, the effect will emerge in future experimental endeavours. The presence of the extra carriers results in an enhancement of the magnitude of the torque between identical slabs. Between non-identical slabs, the sign-reversal effect acts to either increase or decrease the magnitude non-monotonically with separation. We further investigate the influence on the torque by an intervening dielectric medium between the two slabs and find that a medium enhances rather than dampens the interaction for identical slabs. 

Although subject to the limitation of the modeling and computation, the enhancement in the magnitude of the torque predicted is considerable. The numerical computation of the torque using our perturbative formalism proves to be several times faster than numerical computations of Barash's exact result~\cite{Barash1978}, rendering possible comprehensive analyses of the torque for several different cases.

%----------------------------------------
\section{acknowledgement}
We acknowledge support from the Research Council of Norway (Project 221469 and 250346).
 We also acknowledge access to high-performance computing resources via SNIC. PT and JF acknowledge support from the Swedish Research Council. KAM acknowledges support from the US National Science Foundation, grant number 1707511.
%--------------------------------------

%-------------------------------------------------------
\appendix
\begin{widetext}
\section{Reduced reflection coefficients in the presence of an intervening isotropic medium}

We present here the reduced ``reflection coefficients'' 
$\widetilde{\bm R}_i$, for $i=1,2$, for the retarded interaction between planar biaxial materials of finite thickness in the presence of an intervening isotropic fluid. The resulting expressions after separating the dependence on the degree of anisotropy 
$\beta_i$ and relative orientation of the two materials $\theta$ consequent of the polar angle $\phi_k$ integration in the $k$-space, are given by
\begin{equation}\label{reduced-R}
\widetilde{\bm R}_i=
\left[ \begin{array}{lr}
\mathlarger{\frac{1}{\Delta^H}\frac{\kappa_3 \bar\kappa^H_i}{(\kappa_3+\bar\kappa^H_i)^2}} C^H_i
& \mathlarger{\frac{1}{\sqrt{\Delta^E\Delta^H}}\frac{2\kappa_3 \zeta_m\kappa^H_i}
   {(\kappa_3+\bar\kappa^H_i)(\kappa_3+\bar\kappa^E_i)} 
\left(\frac{F_i}{\kappa^E_i+\kappa^H_i}-\frac{G_i}{\kappa^E_i-\kappa^H_i} \right) }
 \\[5mm]
\mathlarger{\frac{1}{\sqrt{\Delta^E\Delta^H}}\frac{2\kappa_3 \zeta_m\kappa^H_i}
   {(\kappa_3+\bar\kappa^H_i)(\kappa_3+\bar\kappa^E_i)} 
\left(\frac{F_i}{\kappa^E_i+\kappa^H_i}-\frac{G_i}{\kappa^E_i-\kappa^H_i} \right)}
& \mathlarger{\frac{1}{\Delta^E}\frac{\kappa_3 \zeta_m^2\varepsilon^{\scriptscriptstyle \perp}_i}{\kappa^E_i(\kappa_3+\kappa^E_i)^2} }C^E_i
\end{array} \right].
\end{equation}
The TE and TM modes do not separate in the case of the 
interaction between anisotropic materials. Thus,
\begin{equation}\label{intgd}
\begin{split}
\textrm{Tr}(\widetilde{\bm R}_1\widetilde{\bm R}_2)=&\quad\frac{1}{(\Delta^H)^2}\frac{\kappa_3^2\bar\kappa^H_1\bar\kappa^H_2}
   {(\kappa_3+\bar\kappa^H_1)^2(\kappa_3+\bar\kappa^H_2)^2}\,C_1^H C_2^H  
+\frac{1}{(\Delta^E)^2}\frac{\kappa_3^2(\zeta_m^2\varepsilon^{\scriptscriptstyle \perp}_1)(\zeta_m^2\varepsilon^{\scriptscriptstyle \perp}_2)}
   {(\kappa_3+\kappa^E_1)^2(\kappa_3+\kappa^E_2)^2 \kappa^E_1\kappa^E_2}\,C_1^E C_2^E  \\
&+ 2 \frac{1}{\Delta^E\Delta^H}\frac{(2\kappa_3)^2\zeta_m^2\kappa^H_1\kappa^H_2}
            {(\kappa_3+\bar\kappa^H_1)(\kappa_3+\kappa^E_1)(\kappa_3+\bar\kappa^H_2)(\kappa_3+\kappa^E_2)}
      \left[\frac{F_1}{\kappa^E_1+\kappa^H_1}-\frac{G_1}{\kappa^E_1-\kappa^H_1} \right]
            \left[\frac{F_2}{\kappa^E_2+\kappa^H_2}-\frac{G_2}{\kappa^E_2-\kappa^H_2} \right],%
\end{split}%
\end{equation}%
%%\twocolumngrid
where $\bar{\Delta}^E= (1 - R_{\overline 1\, \overline 3}^E R_{\overline 2\, \overline 3}^E e^{-2\kappa_3 a})$ and 
$ \bar{\Delta}^H= (1 - R_{\overline 1\,\overline 3}^H R_{\overline 2\,\overline 3}^H e^{-2\kappa_3 a})$. 
The reflection coefficient for the TM mode, for an in-plane isotropic uniaxial dielectric slab, is
\begin{equation}
R_{\overline i\,\overline j}^H 
= - \frac{\bar{\alpha}_{ij}^H(1-e^{-2\kappa_i^H d_i})}{1- (\bar{\alpha}_{ij}^H)^2 e^{-2\kappa_i^H d_i} }, 
 \quad \bar{\alpha}_{ij}^H = \frac{\bar\kappa_i^H-\bar\kappa_j^H}{\bar\kappa_i^H+\bar\kappa_j^H}.
\label{rcoe-magM}
\end{equation}
The TE mode $R_{\overline i\,\overline j}^H$ is obtained by replacing $H\to E$ in
the reflection coefficient for the magnetic mode in
Eq.\,(\ref{rcoe-magM}). Here we have used shorthand notations
\begin{equation}
\kappa_i^H = \sqrt{k^2 \frac{\varepsilon^\perp_i}
 {\varepsilon^{||}_i} +\frac{\zeta_m^2}{c^2}\varepsilon^\perp_i}, 
\qquad
\kappa_i^E = \sqrt{k^2 +\frac{\zeta_m^2}{c^2}\varepsilon^\perp_i}, 
\end{equation}
where $\bar\kappa^H_i = \kappa^H_i/\varepsilon^\perp_i$ and $\bar\kappa^E_i = \kappa^E_i$. For the isotropic intervening medium $\kappa_3^E =\kappa_i^H =\kappa_3=\sqrt{k^2+\frac{\zeta_m^2}{c^2}\varepsilon_3}$ and $\bar\kappa^H_3 = \kappa_3/\varepsilon_3$.
Note the modification of the expressions by the dielectric function $\varepsilon_3$ of the medium as compared to the vacuum case presented in the Supplemental Material of Ref.~\cite{PRL2018}.
We have suppressed the explicit frequency dependence in 
$\varepsilon^{ \perp}$ and $\varepsilon^{||}_i$. The coefficients $C^{H,E}_i$, $F_i$, and $G_i$ are given by
\begin{subequations}
\begin{align}
C_i^X & = \frac{\left\{
          (1-e^{-2\kappa^X_id_i}) \big[1+ (\bar\alpha^X_i)^2 e^{-2\kappa^X_id_i}\big]
     \mp 4\bar\alpha^X_i \kappa^X_id_i e^{-2\kappa^X_id_i}\right\}}
{\big[1-(\bar\alpha^X_i)^2e^{-2\kappa^X_id_i}\big]^2}\xrightarrow{d_i\to\infty} 1
, \label{C-def}\\
F_i &=  \frac{(1-e^{-(\kappa^E_i+\kappa^H_i)d_i})
\left(1- \bar\alpha^E_i\bar\alpha^H_i e^{-(\kappa^E_i+\kappa^H_i)d_i}\right)}
{\left[1-(\bar\alpha^E_i)^2e^{-2\kappa^E_id_i}\right]\left[1-(\bar\alpha^H_i)^2e^{-2\kappa^H_id_i}\right]}\xrightarrow{d_i\to\infty} 1, \\
G_i &=\frac{ \left(\bar\alpha^E_i e^{-\kappa^E_id_i} -\bar\alpha^H_i e^{-\kappa^H_id_i}\right)
\left(e^{-\kappa^E_id_i} - e^{-\kappa^H_id_i}\right)}
{\left[1-(\bar\alpha^E_i)^2e^{-2\kappa^E_id_i}\right]\left[1-(\bar\alpha^H_i)^2e^{-2\kappa^H_id_i}\right]}\xrightarrow{d_i\to\infty} 0.
\end{align}
\end{subequations}
Here $X=H,E$, and the negative and positive signs $(\mp)$ in the numerator  
of Eq.~\eqref{C-def} correspond to $H$ and $E$, respectively. 

\end{widetext}

%--------------
\begin{figure}
\begin{center}
\includegraphics[width=7cm, height=5.5cm]{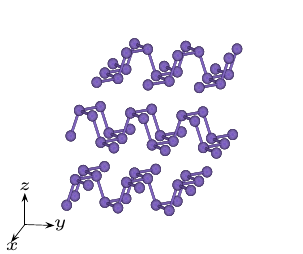}
\caption{(Color online) Structure of black phosphorus indicating the convention of directions followed in this paper.  }
\label{blackP}
\end{center}
\end{figure}
%-------------
%
\section{Structure of black phosphorus}
We show in Fig.~\ref{blackP} the layered structure of black P. In our convention, we define the planar layer by $x$ and $y$ directions while $z$ direction corresponds to the perpendicular direction along the layers. Note, however, that the optical calculation itself is carried out with the bulk primitive unit cell whose three geometrical symmetry axes are not orthogonal to each other. The principal directions are determined by comparing with the orthogonal, conventional unit cell of bulk black P consisting of 8 atoms. Consequently, $x$ is along the direction defined by the lattice parameter 3.34 {\AA}, the so-called zig-zag direction in the literature. $y$ is along the direction spanned by lattice constant 4.47 {\AA} also known as the armchair direction, and $z$ is perpendicular to the layers defined by lattice parameter 10.74 {\AA}. For details of the comparison of the bulk primitive and conventional unit cells that correspond to the optical calculations performed in this work, {\tcc {and further comparison with experimental data}}, see Ref.~\cite{ThiyamThesis}.

%-------------------------------------------------------

\bibliography{achemso}

\end{document}